\documentclass[12pt]{article}
\pdfoutput=1
\usepackage{epsf,amsfonts,amssymb,epsfig,amsmath,mathtools,graphics,slashed}
\usepackage{hyperref,hep,graphicx,subfig}
\usepackage{rotating}
\newcommand{\nn}{\notag \\}

\makeatletter

\@addtoreset{equation}{section}
\makeatother

\begin{document}

\begin{titlepage}

\vfill

\begin{flushright}
Imperial/TP/2015/JG/05\\
DCPT-15/73
\end{flushright}

\vfill

\begin{center}
   \baselineskip=16pt
   {\Large\bf Minimally packed phases in holography}
  \vskip 1.5cm
  \vskip 1.5cm
     Aristomenis Donos$^1$ and Jerome P. Gauntlett$^2$\\
   \vskip .6cm
         \vskip .6cm
      \begin{small}
      \textit{$^1$Centre for Particle Theory\\ and Department of Mathematical Sciences\\Durham University\\Durham, DH1 3LE, U.K.}
        \end{small}\\
        \vskip .6cm
      \begin{small}
      \textit{$^2$Blackett Laboratory, 
        Imperial College\\ London, SW7 2AZ, U.K.}
        \end{small}\\

\end{center}

\vfill

\begin{center}
\textbf{Abstract}
\end{center}
\begin{quote}
We numerically construct asymptotically AdS black brane solutions of $D=4$ Einstein-Maxwell theory coupled to
a pseudoscalar. The solutions are holographically dual to $d=3$ CFTs at finite chemical potential 
and in a constant magnetic field, which spontaneously break translation invariance leading to the spontaneous formation of
abelian and momentum magnetisation currents
flowing around the plaquettes of a periodic Bravais lattice. We analyse the three-dimensional moduli space of lattice solutions,
which are generically oblique, and show, for a specific value of the magnetic field, 
that the free energy is minimised by the triangular lattice, associated with minimal packing 
of circles in the plane. We show that the average stress tensor for the thermodynamically preferred
phase is that of a perfect fluid and that this result applies more generally to spontaneously generated periodic phases.
The triangular structure persists at low temperatures indicating the existence of novel crystalline ground states.
\end{quote}

\vfill

\end{titlepage}

\setcounter{equation}{0}
\section{Introduction}

Spatially modulated phases, in which spatial symmetries are spontaneously broken,
are seen in a wide variety of materials in Nature. For example the cuprate superconductors
exhibit spin and charge density waves as well as current loop order \cite{vojta}. Starting with
\cite{Domokos:2007kt,Nakamura:2009tf} it has become clear that a wide
variety of spatially modulated phases can be realised in strongly coupled field theories using holographic techniques.
These phases are associated with novel black hole solutions in AdS spacetimes which have spatially modulated
horizons\footnote{Sharing some features with the non-holographic solutions of \cite{Wiseman:2002zc}.}, with
the first example presented in \cite{Donos:2012gg}. Exploring the landscape of such black hole
solutions and elucidating their possible zero temperature crystalline ground states is interesting and still largely unexplored territory.

In this paper we continue to investigate a class of models in $D=4$ spacetime dimensions,
which includes the top-down models of \cite{Gauntlett:2009bh},
that couple a metric to a gauge-field and a pseudoscalar field. 
The models all admit an AdS vacuum which is dual to a CFT in $d=3$ spacetime dimensions with
an abelian global symmetry. When the CFT is held at finite chemical potential, $\mu$, with respect to the abelian symmetry, the unbroken spatially homogeneous and
isotropic phase in flat space is dual to the electrically charged AdS-RN black hole (brane). 
Depending on the specific couplings of the model, these black holes can
have striped instabilities
below a critical temperature $T_c$ \cite{Donos:2011bh}. In particular, at $T=T_c$ a static linearised striped mode appears which reveals
the existence of new 
spatially modulated black hole solutions which, furthermore, are associated with abelian and momentum current density waves.
Fully back reacted striped black hole solutions, modulated in one spatial direction and with the current density waves running along the orthogonal spatial direction,
were subsequently constructed in \cite{Rozali:2012es,Donos:2013wia,Withers:2013loa,Withers:2013kva,Rozali:2013ama} by solving a system of PDEs in two variables. 

At the linearised level the static striped modes can be superposed indicating the existence of black holes with spatial
order in two spatial directions. By solving a system of PDEs in three variables, black hole solutions dual to
checkerboard lattices were constructed in \cite{Withers:2014sja}, with the currents now running around the unit cell of the checkerboard. 
It was also shown that the
striped black holes are continuously connected to the checkerboard black holes via rectangular lattice black holes. 
Here we will explain how the checkerboard and rectangular lattice black holes associated with the striped instability of \cite{Donos:2011bh}, 
are special cases of a more general three-dimensional moduli
space of spatially modulated lattice black holes, defined by a two dimensional Bravais lattice. In general these lattices 
are oblique and special cases are rectangular, checkerboard, centred rectangular and triangular lattices.

When there is just a chemical potential, it was shown in \cite{Withers:2014sja}
that the striped phase is thermodynamically preferred over the checkerboard and rectangular lattices (at least for the temperatures and lattices considered in \cite{Withers:2014sja}). In order to obtain two dimensional
phases that are preferred one can supplement $\mu$ with an additional UV deformation. In \cite{Withers:2014sja} it was shown that
after adding a homogeneous source, $\phi_0$, for the operator dual to the pseudoscalar, that the checkerboard phase can be preferred over the striped phase. It would therefore be interesting to further investigate whether other lattices are preferred over the
checkerboard in this context.

In this paper we will not pursue that line of investigation but instead consider the CFT with $\mu\ne0$ and also a uniform
uniform magnetic field\footnote{Spatially modulated phases in the presence of magnetic fields have also been discussed
in \cite{Bolognesi:2010nb,Ammon:2011je,Almuhairi:2011ws,Donos:2011qt,Bu:2012mq,Cremonini:2012ir,Bao:2013fda,Jokela:2014dba}.} of strength $B$. It was shown in \cite{Donos:2012yu} that the striped instability of \cite{Donos:2011bh} can survive when $B\ne 0$.
For a specific value of $B/\mu^2$ we will construct and explore the full three-dimensional moduli space of oblique lattice black holes.
For all these black holes the dual CFT has abelian and momentum magnetisation currents flowing around the unit cells.
The transition from the unbroken phase to the spatially modulated phase is a first
order transition and, interestingly, amongst all of the lattices we find that the triangular lattice is 
the thermodynamically preferred structure. 
The triangular lattice is a commonly observed lattice since it is associated with minimal packing
of circles in the plane. However, why it is preferred in the present context is not so clear given that
we have a strongly coupled plasma. We have cooled the black hole solutions down to quite low temperatures and 
our results indicate the existence of fundamentally new types of holographic crystalline ground state solutions at $T=0$.

In section 2 we introduce the class of models that we study as well as discuss the black hole solutions
that describe the unbroken phase. Section 3 reviews the striped instability while
section 4 discusses a convenient parametrisation of the moduli space of oblique lattices. Section
5 discusses the construction of the lattice black hole solutions and it is shown that the thermodynamically preferred 
solutions are dual to a triangular phase. In carrying out our analysis, 
extending the results of \cite{Donos:2013cka}, we show 
that the average stress tensor is that of a perfect fluid and we argue that this is a general result for spatially modulated phases. Some technical material, including some details on our numerical
implementation, are presented in two appendices.

\section{The setup}
We consider a bulk theory in $D=4$ which couples the metric to a gauge-field, $A$, and a pseudoscalar field, $\phi$, 
with action given by
\begin{align}\label{eq:action}
S=\int\,d^{4}x\,\sqrt{-g}\,&\left(R-V(\phi)-\frac{1}{2}\,\left(\partial\,\phi\right)^{2} -\frac{1}{4}Z(\phi)\,F^{2}\right)\nn&
\qquad\qquad-\frac{1}{8}\,\int\,d^{4}x\,\sqrt{-g}\,\vartheta (\phi)\epsilon^{\mu_{1}\mu_{2}\mu_{3}\mu_{4}}\,F_{\mu_{1}\mu_{2}}F_{\mu_{3}\mu_{4}}\,,
\end{align}
yielding the equations of motion
\begin{align}\label{eq:eom}
R_{\mu\nu}-\frac{1}{2}\partial_{\mu}\phi\,\partial_{\nu}\phi-\frac{1}{2}V\,g_{\mu\nu}+\frac{1}{2}Z\,\left(\frac{1}{4}\,g_{\mu\nu}\,F_{\lambda\rho} F^{\lambda\rho}-F_{\mu\rho}F_{\nu}{}^{\rho}\right)&=0\,,\notag\\
\nabla_{\mu}\left(Z\,F^{\mu\nu} \right)+\frac{1}{2}\partial_{\mu_{1}}\vartheta\,F_{\mu_{2}\mu_{3}}\epsilon^{\mu_{1}\mu_{2}\mu_{3}\nu}&=0\,,\notag\\
\Box\phi-V^{\prime}-\frac{1}{4}Z^{\prime}\,F^{2}-\frac{1}{8}\,\vartheta^{\prime}\,\epsilon^{\mu_{1}\mu_{2}\mu_{3}\mu_{4}}\,F_{\mu_{1}\mu_{2}}F_{\mu_{3}\mu_{4}}&=0\,,
\end{align}
where $F=dA$ is the field strength of the gauge field. In the coordinates we choose later
we take $\epsilon_{trxy}=\sqrt{-g}$. We have taken $16\pi G=1$ for convenience.

The numerical solutions we construct in this paper will
be for specific choices of the functions $V,Z\ge 0,\vartheta$ given in \eqref{models}, below. 
For the moment, though, we just impose some rather weak conditions on these
functions. We demand that $\phi$ is a pseudo-scalar field, 
which is achieved if $V,Z$ are even functions and $\vartheta$ is an odd function of $\phi$:
\begin{align}\label{eq:potentials_sym}
V(-\phi)=V(\phi),\quad Z(-\phi)=Z(\phi),\quad \vartheta(-\phi)=-\vartheta(\phi)\,.
\end{align}
We also assume that the equations of motion
admit an $AdS_4$ vacuum with $\phi=A=0$, which is dual to a $d=3$ CFT with an abelian 
global symmetry and the scalar field is dual to a pseudoscalar operator $\mathcal{O}_\phi$, with scaling dimension $\Delta_\phi$. 
For convenience we will choose $V(0)=-6$ so that the $AdS_4$ vacuum has unit radius.

Our primary focus will be the thermal phase diagram for the CFT with uniform chemical potential $\mu$ and uniform (constant)
magnetic field $B$. The $F\wedge F$ coupling to $\phi$ in the action implies, generically, 
that the scalar field will get activated in the dual black hole solutions. In the case that the scalar is dual to a relevant operator, i.e. $\Delta_\phi<3$, we can also consider
an additional deformation parameter, $\phi_{0}$, associated to sourcing this operator.

At high temperatures, $T/\mu , T/B^{1/2}, T/\phi_{0}^{1/(3-\Delta_{\phi})} >> 1$, the CFT in flat space will be in an unbroken,
homogenous phase which preserves the Euclidean symmetries of the two spatial directions. The associated dual black hole
solutions are captured by a radial ansatz of the form
\begin{align}\label{eq:norm_ansatz}
ds_{4}^{2}&=g^{-2}(r)\,\left(-f(r)\,G(r) dt^{2}+g^{\prime}{}^{2}(r)\,f^{-1}(r)\,G(r)\,dr^{2}+dx^{2}+dy^{2}\right)\,,\notag\\
A&=a_{t}\,dt+\frac{B}{2}\left(x\,dy-y\,dx \right),\quad \phi=\phi(r)\,.
\end{align}
Notice that this does not fully fix the reparametrisation invariance $r\rightarrow R(r)$. For our purposes we found it convenient to choose
\begin{align}
g=r_{+}^{-1}\,(1-(1-r)^{2})\,,
\end{align}
with $r_{+}$ a free constant which sets a scale in the dual field theory. In these coordinates, 
the conformal boundary is at $r=0$. Assuming that $f,G\to 1$ as $r\to 0$, 
writing $r=r_+\epsilon/2$ and taking $\epsilon\to 0$ we find that
\begin{align}\label{bc1}
ds^{2}_{4}\sim \frac{d\epsilon^2}{\epsilon^2}+\frac{1}{\epsilon^2}\,\left( -dt^{2}+dx^{2}+dy^{2}\right)\,.
\end{align}
If we also expand $a_t=\mu+\dots$ and $\phi=\phi_0\epsilon^{3-\Delta_\phi}+\dots$, then we identify
$\mu,B,\phi_0$ as the chemical potential, magnetic field and source for the scalar operator, respectively.

The location of the black hole event horizon is taken to be at $r=1$ and
we impose the following expansions
\begin{align}\label{bc2}
f&=c_{f}\,\left(1-r\right)^{2}+\mathcal{O}\left( (1-r)^{4}\right),\quad G=c_{g}+\mathcal{O}\left((1-r)^{2}\right),\notag\\
a_{t}&=\left(1-r\right)^{2}\,c_{a}+\mathcal{O}\left( (1-r)^{4}\right),\quad \phi=c_{\phi}+\mathcal{O}\left((1-r)^{2}\right)\,.
\end{align}
After the Wick rotation $t \to i\tau$ we find that this leads to analytic behaviour provided that we
periodically identify $\tau $ with period $\beta=\frac{4\,\pi}{r_{+}c_f}$, associated with a temperature $T=\beta^{-1}$.

In the special case that $B=\phi_{0}=0$, the class of theories satisfying \eqref{eq:potentials_sym} admits the exact
electric Reissner Nordstrom black brane solution given\footnote{A more standard version of the AdS-RN solution
is obtained via $r\to 1-(1-r_+/r)^{1/2}$.}
 by
\begin{align}\label{eq:RN}
f(r)&=\frac{1}{4r_{+}^{2}}\,\left(1-r\right)^{2}\,\left( \mu^{2}\,\left( r-2\right)^{3}r^{3}+4r_{+}^{2}\,\left(1+2\,r+3\,r^{2}-4\,r^{3}+r^{4}\right) \right)\,,\notag\\
a_{t}(r)&=\mu\,\left( 1-r\right)^{2},\quad G=1,\quad \phi=0\,,
\end{align}
with temperature $T=\frac{r_{+}}{4\,\pi}\,\left(3-\frac{\mu^{2}}{4r_{+}^{2}}\right)$.
We recall that as $T\to 0$ this solution approaches $AdS_2\times\mathbb{R}^2$ in the far IR.
For general $\mu,B,\phi_0$, the black holes in the ansatz \eqref{eq:norm_ansatz} with boundary conditions
\eqref{bc1},\eqref{bc2} need to be constructed numerically, which can be carried out using standard shooting techniques
(or the techniques outlined in the remaining of the paper).

The specific choices of the functions $V,Z,\vartheta$ that we will focus on in this paper are given by
\begin{align}\label{models}
V=-6\,\cosh\left(\frac{\phi}{\sqrt{3}}\right),\quad Z=\frac{1}{\cosh \left(s\sqrt{3}\,\phi \right)},\quad \vartheta=\chi\,\tanh\left(\sqrt{3}\,\phi \right)\,,
\end{align}
where $(s,\chi)$ are constants. We will quantise the pseudoscalar field so that $\Delta_{\phi}=2$. If
$s=\chi=1$ then this is the top-down model discussed in \cite{Gauntlett:2009bh} (after rescaling the metric, gauge-field
and Newton's constant).

\section{The striped instability}\label{sec:striped_mode}

It was shown in \cite{Donos:2011bh} that when $B=\phi_0=0$ and $\mu\ne 0$, the electric AdS-RN solution \eqref{eq:RN} 
has spatially modulated instabilities below some critical temperature in broad families of theories satisfying \eqref{eq:potentials_sym}. 
The striped instability involves both an electric and momentum current density wave in a direction that is orthogonal to the direction of a spatially modulated 
expectation value for the pseudoscalar operator $\mathcal{O}_\phi$. A simple way to understand the instability is to analyse perturbations
about the $T=0$ $AdS_2\times\mathbb{R}^2$ solution in the far IR, seeking modes with non-trivial dependence on the spatial coordinates of $\mathbb{R}^2$ that violate the $AdS_2$ BF bound.
For the class of models given in \eqref{models} the striped instabilities depend on $(s,\chi)$ as displayed\footnote{Note that $(c_1,n)$ of \cite{Donos:2011bh} equals $(6\sqrt{2}\chi,36 s^2)$. In addition $\tilde m^2_s=-4+36 s^2$.} 
in figure 1 of  \cite{Donos:2011bh} and in particular the top-down model with $s=\chi=1$ is unstable.

When $B$ and $\phi_{0}$ are non-zero, it was shown in \cite{Donos:2012yu}
that there can be a more general modulated instability that also involves modulated charge and energy density waves in addition to
the modulated currents and pseudoscalar. The argument in \cite{Donos:2012yu} for the existence of this instability relied on the fact that the near horizon geometry of the extremal black holes is still $AdS_{2}\times \mathbb{R}^{2}$, which is expected to occur 
in a regime of small deformations, i.e. where $B^{1/2}<<\mu$ and $\phi_{0}^{1/(3-\Delta_{\phi})}/\mu<<1$. However, the existence of such instabilities for generic values of $B^{1/2}/\mu$ and $\phi_{0}^{1/(3-\Delta_{\phi})}/\mu$ is heavily model dependent. In particular, for a given model it
is possible that the $T=0$ ground state of the unbroken phase black holes is no longer $AdS_{2}\times \mathbb{R}^{2}$ 
and moreover that the new IR geometry is perturbatively stable against such modes. An explicit example of this possibility was realised in 
\cite{Withers:2014sja} where it was also shown that, nevertheless, for small enough deformations there can still be
finite temperature spatially modulated instabilities.

We now continue by analysing in more detail the static, perturbative modes about the unbroken phase black hole
solutions \eqref{eq:norm_ansatz} that correspond to the striped instability. Specifically, in a particular gauge we consider
\begin{align}\label{eq:striped_mode}
\delta g_{tt}&=g_{tt}(r)\,h_{tt}(r)\,\cos(\mathbf{k}\cdot\mathbf{x}),\quad \delta g_{rr}=-g_{rr}(r)\,h_{tt}(r)\,\cos(\mathbf{k}\cdot\mathbf{x})\,,\notag\\
\delta g_{ti}&=\hat{n}_{i}\,h_{t\perp}(r)\,\sin(\mathbf{k}\cdot\mathbf{x}),\notag\\
\delta g_{ij}&=\left(\delta_{ij}-\hat{k}_{i}\hat{k}_{j}\right)\,h_{\perp\perp}(r)\,\cos(\mathbf{k}\cdot\mathbf{x})+\hat{k}_{i}\hat{k}_{j}\,h_{\parallel\parallel}(r)\,\cos(\mathbf{k}\cdot\mathbf{x})\,,\notag\\
\delta a_{t}&= h_{t}(r)\,\cos(\mathbf{k}\cdot\mathbf{x}),\qquad\qquad \delta a_{i}=\hat{n}_{i}\,h_{\perp}(r)\,\sin(\mathbf{k}\cdot\mathbf{x}),\notag\\
\delta \phi&=h(r)\,\cos(\mathbf{k}\cdot\mathbf{x}),
\end{align}
where the indices $i,j$ run over the spatial coordinates $(x,y)$ and
\begin{align}
\hat{k}&=\mathbf{k}/\|\mathbf{k}\|,\quad \hat{n}\cdot\hat{k}=0,\quad \|\hat{n}\|=1\,.
\end{align}
In addition $g_{tt}$ and $g_{rr}$ are the background metric components in \eqref{eq:norm_ansatz}. 
By adding this perturbation to the background \eqref{eq:norm_ansatz} and substituting into the equations of
motion for \eqref{eq:action}, at linear order we are led to a system of second order ODEs for $\left\{h_{\perp\perp},h_{t\perp},h_{t},h_{\perp},h \right\}$ and first order ODEs for $\left\{ h_{tt}, h_{\parallel\parallel}\right\}$. A solution is therefore
specified by twelve integration constants.
For the case of 
the AdS-RN black hole solution \eqref{eq:RN} (with $B=\phi=0$), we go back to the case studied in \cite{Donos:2011bh} with only the set
$\left\{h_{t\perp},h_{\perp},h \right\}$ being non-trivial. 

Demanding regularity of the perturbation close to the horizon gives the analytic expansion at $r=1$:
\begin{align}
h_{\mu\mu}&=c_{\mu\mu}+\mathcal{O}\left( (1-r)^{2}\right),\qquad\qquad h_{t\perp}=c_{t\perp}\,\left(1-r\right)^{2}+\mathcal{O}\left( (1-r)^{4}\right),\notag\\
h_{t}&=c_{t}\,\left(1-r\right)^{2}+\mathcal{O}\left((1-r)^{4}\right),\quad h_{\perp}=c_{\perp}+\mathcal{O}\left( (1-r)^{2}\right),\notag\\
h&=c+\mathcal{O}\left( (1-r)^{2}\right)\,,
\end{align}
where $\left\{c_{\mu\nu},c_{ty},c_{\mu},c\right\}$ are constants of integration to be determined.
In the absence of sources, with $\Delta_\phi=2$, the corresponding UV expansion near $r=0$ is given by
\begin{align}
h_{\mu\nu}=d_{\mu\nu}\,r+\mathcal{O}(r^{2}),\quad h_{\mu}=d_{\mu}\,r+\mathcal{O}(r^{2}),\quad h=d\,r^{2}+\mathcal{O}(r^{3})\,,
\end{align}
where $\left\{d_{\mu\nu},d_{\mu},d\right\}$ are additional constants,
out of which $d_{tt}$ and $d_{xx}$ are fixed in terms of the others. In total we have seven constants from the near horizon expansion and another five from the asymptotic region. Since we are dealing with a linear system of equations, we can scale the unknown functions to set any of the twelve non-zero constants to one. The extra constant needed to uniquely specify a solution for fixed values of $\mu$, $\|\mathbf{k}\|$, $B$ and $\phi_{0}$ is the critical temperature, $T_c/\mu$, at which the static mode comes into existence, and below
which the black holes are unstable.

\begin{figure}[h!]
\centering
\includegraphics[width=0.5\textwidth]{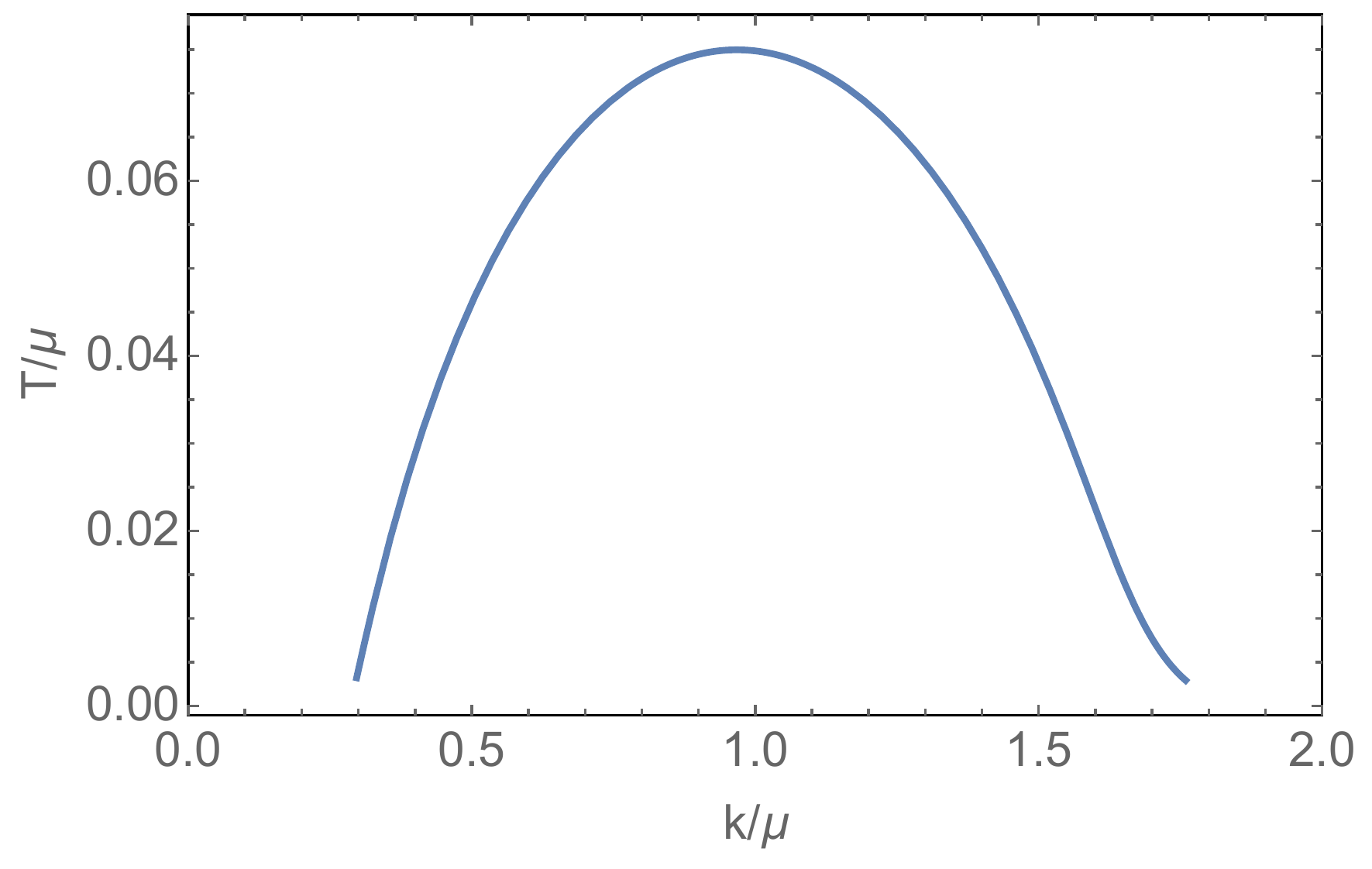}
\caption{For black holes in the unbroken phase with $B/\mu^2=0.05$ we plot the critical temperature, $T_c$,
for the onset of a striped instability with wave number $k$. 
The bulk model is as in \eqref{models} with $(s,\chi)=(1,1.5)$.
The maximum temperature for which we find a zero mode is at $T/\mu=0.075$ with $k/\mu=0.967$.
\label{bell}}
\end{figure}

We mentioned above that for the electric AdS-RN black hole the top down model with $s=\chi=1$ is unstable. 
For a given value of $s$, the critical temperature for the instability is an increasing function of $\chi$. For $s=1$, 
the top-down value $\chi=1$ is very close to a critical value $\chi_{c}$ below which the unstable modes completely disappear. In the rest of the paper we will therefore take $s=1$ and $\chi=1.5$ in order to improve the
numerics, but we expect that the results for the top-down values will be similar.
Setting the scalar deformation to zero, $\phi_0=0$, we have numerically constructed black holes dual
to the unbroken phase with $\mu\ne 0$ and $B/\mu^2=0.05$. One finds that the $T=0$ limit of these black holes 
approaches a dyonic $AdS_2\times\mathbb{R}^2$ solution in the IR and hence, using the results
of \cite{Donos:2012yu}, has striped instabilities. We also constructed the critical temperatures at which the
striped instability appears and the results are presented in figure \ref{bell}.

\section{The moduli space of oblique lattices}

Before moving on to the discussion of the back reacted solutions that are associated with the striped modes given in
\eqref{eq:striped_mode} we first discuss the moduli space we wish to explore. It is clear that at the critical
temperature $T_c$, associated with a fixed value of the magnitude
of the wave-number, $k=\|k\|$,  there is a one parameter
family of such 
striped modes, related by a rotation in the $(x,y)$ plane. At the linearised level we can consider an arbitrary
linear superposition of these modes. However, demanding periodicity in the $(x,y)$ plane, so that the variational problem for the back reacted solutions is well-posed, restricts the superposition to be the sum of two wave-number vectors, 
each with magnitude $k$ and in general non-parallel. Thus, the resulting set of zero modes at $T_c$ 
is parametrised by an angle between the two vectors.

By including the back reaction we therefore expect a family of periodic spatially modulated solutions to appear at $T_c$.
For lower temperatures the moduli space of periodic black hole solutions will be parametrised by two wave-numbers
$\mathbf{k}_{1}$ and $\mathbf{k}_{2}$, which no longer need to have the same magnitude. In other words we allow
the spatial periodicities to change as we lower the temperature.
It is always possible to perform a rotation to choose, for example, $\mathbf{k}_{1}$ to be parallel to the $y$-direction. 
This leaves us with a three parameter moduli space of black hole solutions to explore. 
In the special case that the $\mathbf{k}_{i}$ are parallel, with same magnitude,  
we have stripes, otherwise we will
have a two-dimensional Bravais lattice. Generically these are oblique periodic lattices, but there are various special cases: rectangular lattices, checkerboards, centred rectangular lattices and hexagonal lattices. The latter is also
known as a triangular lattice and is the lattice associated with
minimal packing of circles in the plane. The task is to explore this three-dimensional moduli space of
black holes, and identify the configuration that minimises the free energy at each temperature $T\le T_c$.

It is clear from this discussion that the back-reacted striped black hole solutions constructed in 
\cite{Donos:2013wia,Withers:2013loa,Withers:2013kva,Rozali:2012es,Rozali:2013ama} have additional zero modes associated with the larger moduli space of solutions corresponding to the two-dimensional lattices.
For the special case of $\mu\ne 0$, $\phi_0= 0$ and $B=0$, checkerboard lattices were constructed
in \cite{Withers:2014sja} but were shown to have free energy that is sub-dominant to the striped phase. Some additional
rectangular lattices were also constructed in \cite{Withers:2014sja} for this case and also shown to be sub-dominant.
As an aside, as far as this paper is concerned, we note that some preliminary investigations which we have carried out indicate that the striped phase is also
preferred over all oblique lattices.
It was also shown in \cite{Withers:2014sja} that when $\mu\ne 0$, $\phi_0\ne 0$ and $B=0$, the checkerboards
can dominate over the stripes and the transition is first order. 
It remains an interesting open problem to revisit the analysis of \cite{Withers:2014sja} and explore
the full three-dimensional moduli space of periodic lattices when $\phi_0\ne 0$ and $B=0$.

In the next section we instead construct back reacted black hole solutions with $\mu\ne 0$, $\phi_0= 0$ and $B\ne 0$ and explore the full moduli space of solutions. Before doing that we briefly discuss a convenient 
parametrisation of the periodic configurations. Let us now denote the spatial coordinates $(x',y')$
associated with the asymptotic boundary metric
\begin{align}\label{eq:asympt}
ds_{3}^{2}=-dt^{2}+(dx')^{2}+(dy')^{2}\,.
\end{align}
Consider wave numbers
 \begin{align}
\mathbf{k}_{x}=\frac{2\pi}{L_{x}}\,\left(\sin\alpha,\cos\alpha \right),\quad\mathbf{k}_{y}=\frac{2\pi}{L_{y}}\,\left(0,1 \right) \,,
\end{align}
with $\pi/2\le\alpha\le\pi$. The associated periodic boundary conditions in the $(x',y')$ directions are given by
\begin{align}\label{pcsone}
\left(x^{\prime},y^{\prime}\right)\equiv \left(x^{\prime}+\frac{L_{x}}{\sin\alpha}\,n_{1}-\cot \alpha\,L_{y}\,n_{2},y^{\prime}+L_{y}\,n_{2} \right)\,,
\end{align}
where $n_i$ are integers.
The parameters $L_{x}$, $L_{y}$ and $\alpha$ are the three moduli for the space of solutions that
we would like to construct. 
We also note that in the plots involving the spatial coordinates that we present later we use $(\hat x,\hat y)\equiv (\frac{x'}{L_x},\frac{y'}{L_y})$.

An equivalent parametrisation, which we shall use in the next section, is to impose the simple periodic boundary conditions
\begin{align}\label{eq:periodicity}
\left(x,y\right)\equiv \left(x+n_1,y+n_2 \right)\,,
\end{align}
but demand that the conformal metric asymptotes to
\begin{align}\label{eq:asympt2}
ds_{3}^{2}=-dt^{2}+\frac{1}{\sin^{2}\alpha}\,\left(L_{x}^{2}\,dx^{2}+L_{y}^{2}\,dy^{2}-2\,L_{x}\,L_{y}\,\cos\alpha\,dx\,dy \right)\,.
\end{align}
Indeed the two coordinate systems are related by
\begin{align}
x=L_{x}^{-1}\,\left(\cos\alpha\,y^{\prime}+\sin\alpha\,x^{\prime}\right),\quad y&=L_{y}^{-1}\,y^{\prime} \,.
\end{align}

Instead of $(L_{x},L_{y},\alpha)$ we can also parametrise the moduli space of solutions using $(k_x,k_y, R_0)$ with $k_x\equiv 2\pi/L_x$, $k_y\equiv 2\pi/L_y$, $R_0\equiv -\cot\alpha$ and $R_0\in[0,\infty)$. Notice that $R_0=0$ 
($\alpha=\pi/2$) corresponds to a rectangular lattice and when $k_1=k_2$ it becomes a checkerboard. As $R_0\to\infty$ 
($\alpha=\pi$) we obtain stripes. The hexagonal, or triangular lattice, is given by $k_1=k_2$ and $R_0=1/\sqrt{3}$ ($\alpha=2\pi/3$).

All lattices can be obtained by allowing $\alpha$ to lie in the range $\pi/2\le \alpha\le \pi$ but there is
some redundancy. Generically we have an oblique lattice. There are the following special cases with larger symmetry.
If $k_x=k_y$ we have rhombic lattices; if in addition $\alpha=2\pi/3$ it is hexagonal while if $\alpha=\pi/2$ it is square.
If $\alpha=\pi/2$ we have rectangular lattices; if in addition $k_x=k_y$ we have a square lattice.
If $k_y/k_x=-2\cos\alpha$ we have centred rectangular lattices satisfying $|\mathbf{k}_{y}|^2=-2\mathbf{k}_{y}\cdot\mathbf{k}_{x}$; if in addition $\alpha=2\pi/3$ it is the hexagonal lattice, while if $\alpha=3\pi/4$ (i.e. $k_y=\sqrt{2}k_x$) 
it is the square lattice rotated by $\pi /4$ compared to the previous two. Finally there are also similar
centred rectangular lattices with $k_x/k_y=-2\cos\alpha$.

\section{Lattice black holes solutions}\label{sec:non-linear}

We are now in the position to numerically construct the back reacted black hole solutions for the periodic 
lattices that we described in the previous section. To achieve this, we use the following
ansatz which has only a time-like Killing vector:
\begin{align}\label{eq:pde_ansatz}
ds_{4}^{2}&=g^{-2}(r)\left(-\hat{f}(r)\,Q_{tt}\, (\eta^{t})^{2}+\frac{g^{\prime}{}^{2}(r)}{\hat{f}(r)}\,Q_{rr}\,\,dr^{2}+Q\,\sqrt{1+R^{2}}\,\left[M\,(\eta^{x})^{2}+\frac{1}{M}\,(\eta^{y})^{2} \right]+2\,Q\,R\,\eta^{x}\eta^{y}\right),\notag\\
\eta^{t}&=dt+\hat{L}_{x}\,Q_{tx}\,dx+\hat{L}_{y}\,Q_{ty}dy+Q_{tr}dr\,,\notag\\
\eta^{x}&=\hat{L}_{x}\,dx+Q_{rx}\,dr,\quad \eta^{y}=\hat{L}_{y}\,dy+Q_{ry}\,dr\,,\notag\\
A&=\frac{1}{4}\,g^{\prime}(r)^{2}\,a_{t}\,dt+\frac{g'(r)}{g(r)}a_{r}\,dr+\hat{L}_{x}a_{x}\,dx+\hat{L}_{y}a_{y}\,dy+\hat{L}_{x}\,\hat{L}_{y}\frac{B}{2}\left(x\,dy-y\,dx \right)
%\,B\,dx\wedge dy
\,,\notag\\
\phi&=g(r)\,\varphi\,,
\end{align}
where $\{Q_{tt}, Q_{rr}, Q, R, M, Q_{tx}, Q_{ty}, Q_{tr}, Q_{rx}, Q_{ry}\}$ and
$\{a_t, a_r, a_x, a_y, \varphi\}$ are all functions of $r$ and periodic functions of $\mathbf{x}=(x,y)$ with
$(x,y)\sim (x+1,y)\sim (x,y+1)$.
The functions $\hat{f}(r)$ and $g(r)$ that appear in the ansatz \eqref{eq:pde_ansatz} are fixed by hand for convenience (c.f. \eqref{eq:RN}) and their explicit form is taken to be
\begin{align}
\hat{f}&=\frac{1}{4r_{+}^{4}}\,\left(1-r\right)^{2}\,\left( \left(\mu^{2}\,r_{+}^{2}+B^{2}\right)\,\left( r-2\right)^{3}r^{3}+4r_{+}^{4}\,\left(1+2\,r+3\,r^{2}-4\,r^{3}+r^{4}\right) \right)\,,\notag\\
g&=r_{+}^{-1}\,(1-(1-r)^{2})\,.
\end{align}
For convenience we will choose
\begin{align}\label{conchoice}
\hat{L}_{x}=\frac{L_{x}}{\sqrt{\sin\alpha}},\quad \hat{L}_{y}=\frac{L_{y}}{\sqrt{\sin\alpha}}\,.
\end{align}
Of course, there is a lot of redundancy in this ansatz due to local coordinate and gauge invariance of the theory \eqref{eq:action}
and this will be dealt with below.

The boundary conditions near the $AdS_{4}$ boundary, located at $r=0$, that we want to impose are given by
\begin{align}
Q_{tt}(0,\mathbf{x})&=Q_{rr}(0,\mathbf{x})=Q(0,\mathbf{x})=M(0,\mathbf{x})=1,\quad R=R_0\equiv-\cot\alpha,\notag\\
Q_{rx}(0,\mathbf{x})&=Q_{ry}(0,\mathbf{x})=Q_{tr}(0,\mathbf{x})=Q_{tx}(0,\mathbf{x})=Q_{ty}(0,\mathbf{x})=0,\notag\\
a_{t}(0,\mathbf{x})&=\mu,\quad a_{r}(0,\mathbf{x})=a_{x}(0,\mathbf{x})=a_{y}(0,\mathbf{x})=0,\notag\\
\varphi(0,\mathbf{x})&=\phi_{0}\,,
\end{align}
and we will only consider $\phi_{0}=0$, the case of no scalar deformation. Notice that these boundary conditions
imply that the asymptotic metric approaches \eqref{eq:asympt} and we therefore demand that the coordinates $x$ and $y$ 
have unit periods, as in \eqref{eq:periodicity}.
Also observe that the $B$ dependent part of the field strength for the gauge field takes the form
$\hat L_x \hat L_yBdx\wedge dy=Bdx'\wedge dy'$ in the coordinates associated with
\eqref{eq:asympt} and \eqref{pcsone}.

We will demand that there is an analytic Killing horizon generated by the Killing vector $\partial_t$, located
at $r=1$. Analyticity demands that our functions will only depend on even powers of $(1-r)$, leading to the Neumann boundary condition $\partial_{r}\Phi(1,x,y)=0$ for all functions appearing in the ansatz \eqref{eq:pde_ansatz} apart from $Q_{tt}$, for which we impose $Q_{tt}\left(1,x,y \right)=Q_{rr}\left(1,x,y\right)$.

In order to get a well defined boundary valued problem for an elliptic, quasi-linear system of partial differential
equations, as well as dealing with the residual coordinate and gauge invariance,
we follow the approach of \cite{Headrick:2009pv,Figueras:2011va,Adam:2011dn,Wiseman:2011by}
by suitably modifying the equations of motion. After solving the modified equations, we check that the solutions
in fact solve the unmodified equations.

For the Einstein equations we add the term $\nabla_{(\mu}\xi_{\nu)}$ to the left hand side of Einstein's equation in \eqref{eq:eom}. 
As in \cite{Headrick:2009pv} we choose $\xi^{\mu}=g^{\nu\lambda}\left(\Gamma^{\mu}_{\nu\lambda}-\bar{\Gamma}^{\mu}_{\nu\lambda} \right)$ where the Christoffel symbol $\bar{\Gamma}^{\mu}_{\nu\lambda}$ is with respect to a reference metric $\bar{g}_{\mu\nu}$ which needs to have a Killing horizon and the same asymptotics with the solution we would like to construct.
In the case of pure gravity and non-positive cosmological constant, one can actually show \cite {Figueras:2011va} that given such a vector, there is no non-trivial solutions to the modified Einstein's equation with $\xi^{\mu}$ non-trivial.  
For the more general kind of theories we are considering, though, a corresponding theorem is not yet available and one has to numerically check that $\xi^{\mu}=0$ a posteriori. The vanishing of $\xi^\mu$ fixes the choice of coordinates.

Following a similar logic for the gauge field equation of motion, we will introduce a scalar $\psi$ and, as in \cite{Withers:2014sja}, 
add the  term $Z\,\nabla^{\nu}\psi$ to the left hand side of Maxwell's equation in \eqref{eq:eom}. With the choice of $\xi^\mu$ above, we choose\footnote{This differs from the choice made in 
\cite{Withers:2014sja}, which in general does not lead to elliptic PDEs.} 
$\psi$ to be given by
\begin{align}\label{defpsi}
\psi&=\nabla_{\mu}\left(A^{\mu}-\bar{A}^{\mu} \right)+\xi^{\mu}\left(A_{\mu}-\bar{A}_{\mu} \right)\,,\nn 
&=\nabla_{\mu}A^{\mu}+\xi_{\mu}A^{\mu}-g^{\mu\nu}\bar\nabla_\mu\bar{A}_{\nu} \,.
\end{align}
By linearising the metric and the gauge-field about a given configuration one can check that this modification
leads to an elliptic set of equations. In fact this is true whether or not we include the terms involving $\bar A$:
the utility of these terms is that they allow one to choose $A^{\mu}-\bar{A}^{\mu}$ conveniently.
In the present set up we will choose $\bar A=\hat{L}_{x}\,\hat{L}_{y}\frac{B}{2}\left(x\,dy-y\,dx \right)$ so that
the components of $A^{\mu}-\bar{A}^{\mu}$ are periodic functions and so that $\psi=0$ on the boundary.
We will also choose the reference metric
$\bar{g}_{\mu\nu}$ to be given by \eqref{eq:pde_ansatz} with 
$Q_{tt}=Q_{rr}=Q=M=1$, $R=R_0$, 
%$a_t=\mu$ 
and vanishing
 $\{Q_{tx},Q_{ty},Q_{tr},Q_{rx},Q_{ry},\}$. Notice that $\psi$ is now a periodic function of 
 $\mathbf{x}$ which vanishes at the $AdS$ boundary and satisfies Neumann boundary conditions
 at the black hole horizon.

If we take the covariant derivative of the modified Maxwell equations we deduce that
\begin{align}
\nabla^{\mu}\left(Z\,\nabla_{\mu}\psi\right)=0\,.
\end{align}
We now multiply this equation by $\psi$ and integrate over $r,{\bf x}$. Integrating by parts we find
that the boundary contributions vanish and, with $Z\ge0$, we conclude $\nabla\psi=0$. Since $\psi=0$
on the AdS boundary we conclude that $\psi=0$ everywhere. Similarly to the metric, vanishing of $\psi$ fixes the gauge invariance. 

To summarise, if we solve our modified equations we just need to check that $\xi^\mu=0$ to ensure
that we have solved the equations of motion \eqref{eq:eom}. Since $\xi^\mu$ is spacelike, this is achieved by checking $\xi^2=0$.
Some details about the asymptotic expansion for the modified equations are included in appendix \ref{app:expansion} and some discussion of the numerical implementation and convergence can be found in appendix \ref{app:reference}.

\subsection{Currents and free energy}
We can extract the current and stress tensor of the dual field theory from the
asymptotic behaviour of the metric and gauge-field. As shown in appendix \ref{app:expansion}, and following
\cite{Donos:2015bxe}, the spatial components of the abelian and momentum currents are magnetisation currents
of the form
\begin{align}\label{defjays}
\sqrt{\gamma}J^i=\partial_j M^{ij}\,,\qquad
\sqrt{\gamma}(T^{ti}-\mu J^i)&=\partial_j M^{ij}_T\,,
\end{align}
where $M^{ij}=M^{[ij]}$ is the local magnetisation density
and $M^{ij}_T=M^{[ij]}_T$ is the local
thermal magnetisation density.
This immediately implies that the average current fluxes vanish $\bar J^{i}=\bar T^{ti}=0$, where here,
and in the following, the bar refers to a period average e.g.
\begin{align}
\bar J^i&=\frac{\sin\alpha}{L_xL_y}\int d^2x'\sqrt{\gamma}J^i\,.
\end{align}
Here we are using the periodic coordinates given in \eqref{pcsone},
$\sqrt{\gamma}$ refers to the flat spatial part of the metric, $\gamma_{ij}$, 
in \eqref{eq:asympt} and we have used the fact that the 
volume of the unit cell on the torus is given by $\sin\alpha/(L_xL_y)$. 
In all of the black hole lattice solutions that we construct
there are both abelian and momentum currents flowing around the unit cells as in figure \ref{honey}.

In order to calculate the thermodynamically preferred black hole solutions we need to determine the free energy.
The
free energy density is given by
\begin{align}
w=\frac{\sin\alpha}{L_xL_y}TI_{OS}\,,
\end{align}
where $I_{OS}$ is the total Euclidean on-shell action, including contributions from the Gibbons-Hawking
terms as well as standard counter terms. We note that for the black holes of interest we have $w=w(T,\mu,B;k_1,k_2,\alpha)$.
For a fixed type of periodic lattice. i.e. fixing $(k_1,k_2,\alpha)$, we have
\begin{align}\label{varfe}
\delta w=-s\delta T-\bar J^t\delta \mu-m\delta B\,,
\end{align}
where $s$ is the entropy density (i.e. entropy per unit cell area), 
$\bar J^t$ is the average charge density and 
$m$ is the magnetisation density given by the bulk integral, as in \cite{Donos:2012yu},
\begin{align}\label{defemm}
m=-\frac{\sin\alpha}{L_xL_y}\int dr dx' dy'\left(\sqrt{-g}F^{x'y'}Z-\vartheta F_{tr}\right)\,,
\end{align}
In appendix \ref{app:expansion} we derive an expression for the local magnetisation density, $M^{ij}$,
which, consistent with \eqref{defjays}, has an ambiguity of the addition of a constant times $\epsilon^{ij}$. 
This constant can be fixed, thermodynamically, by demanding that $m=\bar M^{x'y'}$.

If we instead hold fixed the UV data $(T,\mu,B)$ and vary $(k_1,k_2,\alpha)$, by following
the arguments of \cite{Donos:2013cka}, we find after switching from $(x',y')$ to $(x,y)$ coordinates
\begin{align}\label{dertran}
k_1\frac{\delta w}{\delta k_1}&=w+mB+\bar T^{x}{}_{x}\,,\nn
k_2\frac{\delta w}{\delta k_2}&=w+mB+\bar T^{y}{}_{y}\,,\nn
\frac{\delta w}{\delta \alpha}&=\cot\alpha (w+mB+\bar T^{x}{}_{x}+\bar T^{y}{}_{y})-\frac{L_xL_y}{\sin\alpha}
 \bar T^{xy}\,,
\end{align}
where $\bar T$ is the period average of the stress tensor. In addition, we also have
\begin{align}
w=-sT-\bar J^t\mu +\bar T^{tt}\,.
\end{align}
For the thermodynamically preferred black hole solutions, all of the derivatives in \eqref{dertran} vanish, and it
is straightforward to show that
we have 
\begin{align}\label{extreme}
w&=-(mB+p)\,,\nn
\bar T^{ij}&=p\gamma^{ij}\,,\end{align}
where $\gamma^{ij}$ is the spatial part of the
flat metric given in the $(x,y)$ coordinates in \eqref{eq:asympt} and the constant $p$ is the average pressure.
We mentioned earlier that $\bar T^{ti}=0$. It is interesting to note that in spite of the spatial modulation the average stress tensor for the thermodynamically preferred
phase is that of a perfect fluid\footnote{Although we obtained this result for
spatially modulated phases in two flat spatial dimensions, it is clear that it generalises.
Specifically, in $d$ spatial dimensions there will be $d$ conditions analogous to the first two lines in
\eqref{dertran} and an additional $d(d-1)/2$ conditions for pairs of coordinates analogous
to the last line in \eqref{dertran}. This will then lead to \eqref{extreme}.}. 
We also notice that for the thermodynamically preferred black holes we have
$sT+\bar J^t\mu=\bar T^{tt}+(mB+p)$.

\subsection{Results}
With $B/\mu^2=0.05$ we recall that the critical temperature for the onset of the instability is given by $T/\mu=0.075$ (see figure \ref{bell}). We have constructed various oblique lattice black holes for temperatures both greater than and less than
$T/\mu=0.075$. Since some oblique lattice black holes exist for $T/\mu>0.075$ which are not thermodynamically preferred
and since they are all thermodynamically preferred for 
$T/\mu<0.075$, it is clear that the transition from the homogeneous phase to the spatially modulated phase is a first order 
transition. 

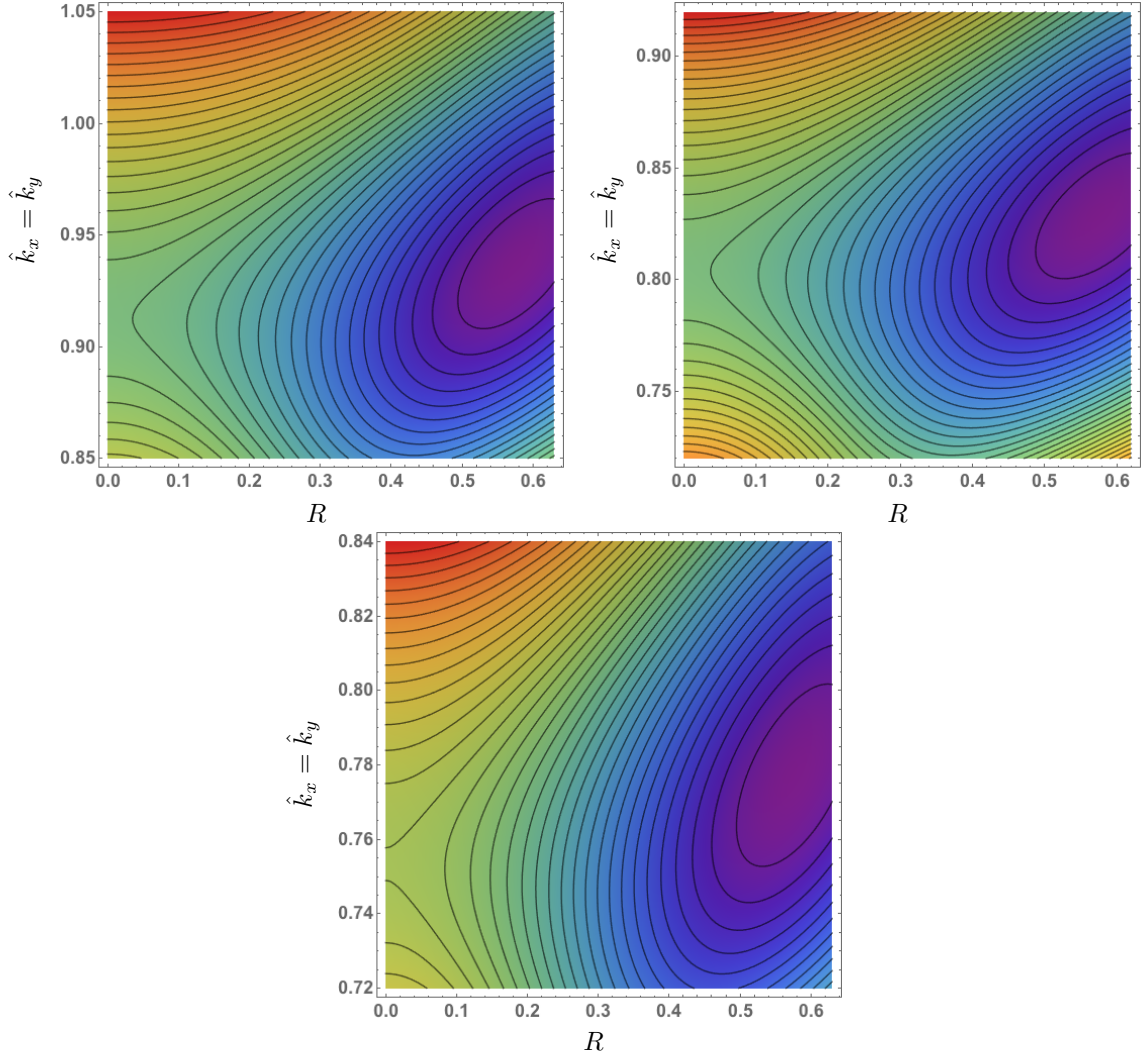
\begin{figure}[h!]
\centering
\includegraphics[width=0.45\textwidth]{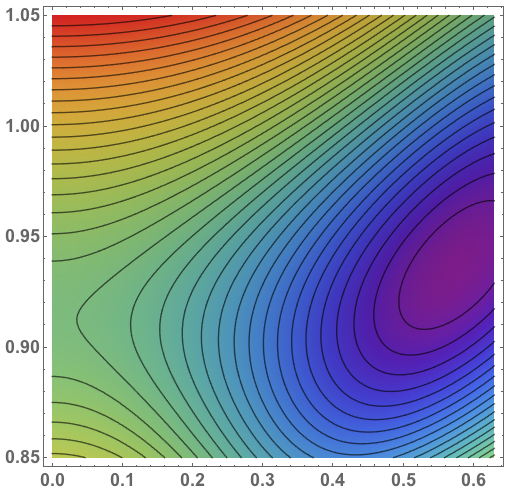}\qquad
\includegraphics[width=0.45\textwidth]{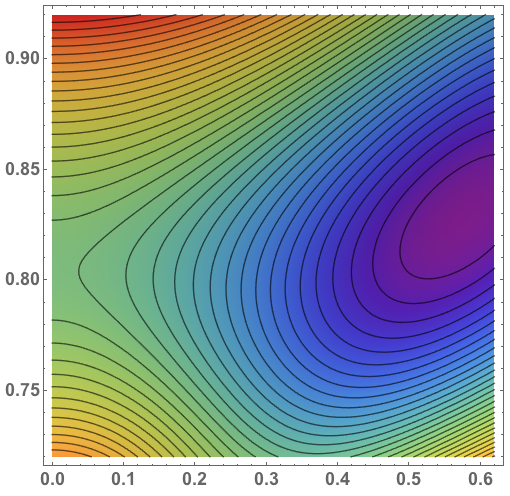}\\
\vspace{10pt}
\includegraphics[width=0.45\textwidth]{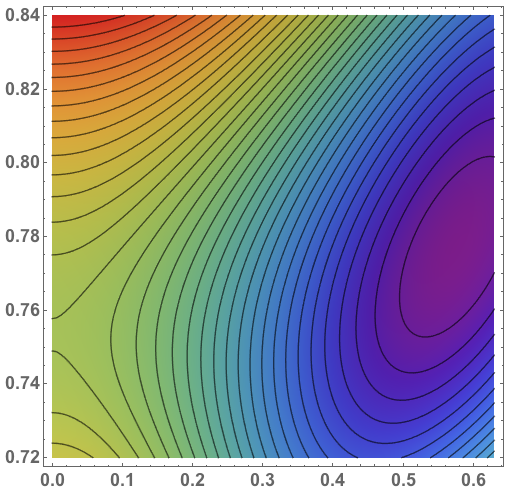}
\begin{picture}(0.1,0.1)(0,0)
\put(-95,306){\makebox(0,0){\footnotesize{\begin{sideways}$\hat{k}_{x}=\hat{k}_{y}$\end{sideways}}}}
\put(-315,306){\makebox(0,0){\footnotesize{\begin{sideways}$\hat{k}_{x}=\hat{k}_{y}$\end{sideways}}}}
\put(-210,100){\makebox(0,0){\footnotesize{\begin{sideways}$\hat{k}_{x}=\hat{k}_{y}$\end{sideways}}}}
\put(15,195){\makebox(0,0){\footnotesize{$R$}}}
\put(-205,195){\makebox(0,0){\footnotesize{$R$}}}
\put(-100,-5){\makebox(0,0){\footnotesize{$R$}}}
\end{picture}
\vspace{5pt}
\caption{Free energy of the rhombiboard lattices. The contours display the free energy density $\hat{w}\equiv w/\mu^{3}$ as a function of
$R_0$ and $\hat{k}_{x}=\hat{k}_{y}$, where $\hat{k}_{x}\equiv k_{x}/\mu$ and $\hat{k}_{y}=k_{y}/\mu$, for temperatures $T/\mu=0.07$ (top left), $T/\mu=0.05$ (top right) and $T/\mu=0.04$ (bottom). 
Colours with longer wavelength correspond to contours with larger values.
The minimum is at $R_0=1/\sqrt{3}$, corresponding to triangular lattices, 
in all three cases but the period changes with temperature. The minima for the three cases are located at $\hat{k}_x=0.937$, $\hat{k}_x=0.828$ and $\hat{k}_x=0.777$, respectively.
\label{fig:fingerprint}}
\end{figure}

Our principal interest is not to investigate the details of the first order transition (for example the critical temperature), but
rather to investigate the shape of the thermodynamically preferred lattices and so we just discuss
solutions with $T/\mu<0.075$. In order to explore the moduli space of black hole solutions, we first consider rhombic lattices (``rhombiboards") for which $k_x=k_y$ (i.e. $L_x=L_y)$. We present some results for the
black hole solutions that we constructed for the three different temperatures, 
$T/\mu=0.07, 0.05$ and $0.04$. In the ``dactylograms" of figure \ref{fig:fingerprint}, each of which was made from about $400$
individual black hole solutions, we show the contours of the
free energy as a function of $k_x$ and $R_0$.

In all three cases, quite remarkably, we find that the 
minimum is given\footnote{The minimum can be obtained in two ways. One can construct the free 
energy, stress tensor and magnetisation for the individual black holes and then use \eqref{dertran} to 
construct the derivatives on the left hand side of \eqref{dertran} as functions of $(k_x,\alpha)$ via an interpolation. 
The vanishing of these derivatives gives $R_0=1/\sqrt{3}$ to about one part in $10^7$. An alternative method 
is simply to use the free energy density for the individual black holes to
obtain a function $w$ of $(k_x,\alpha)$ via an interpolation. Setting the derivatives of this function to zero
gives $R_0=1/\sqrt{3}$ to about one part in $10^4$.} by $R_0=1/\sqrt{3}$ with the value of $k_x$ decreasing as the temperature is lowered. 
In other words at all three temperatures the triangular lattice are the preferred lattices within the class of
rhombic lattices. As an aside we notice in figure \ref{fig:fingerprint} that for each case there is a saddle point at $R_0=0$, associated with checkerboards.

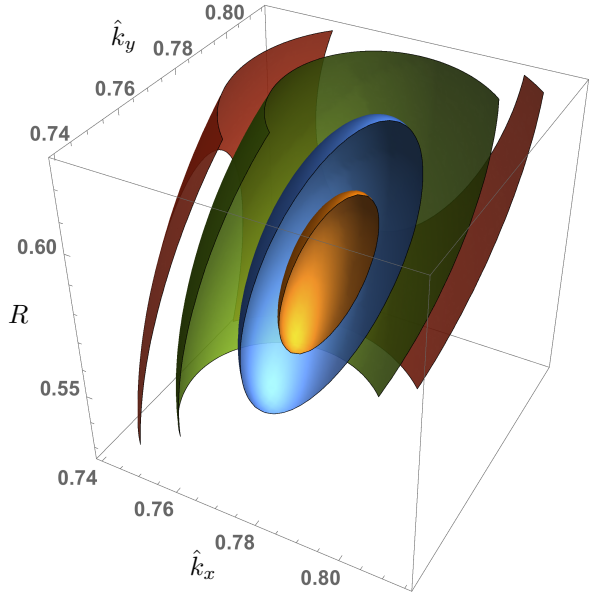
\begin{figure}[t!]
\centering
\includegraphics[width=0.5\textwidth]{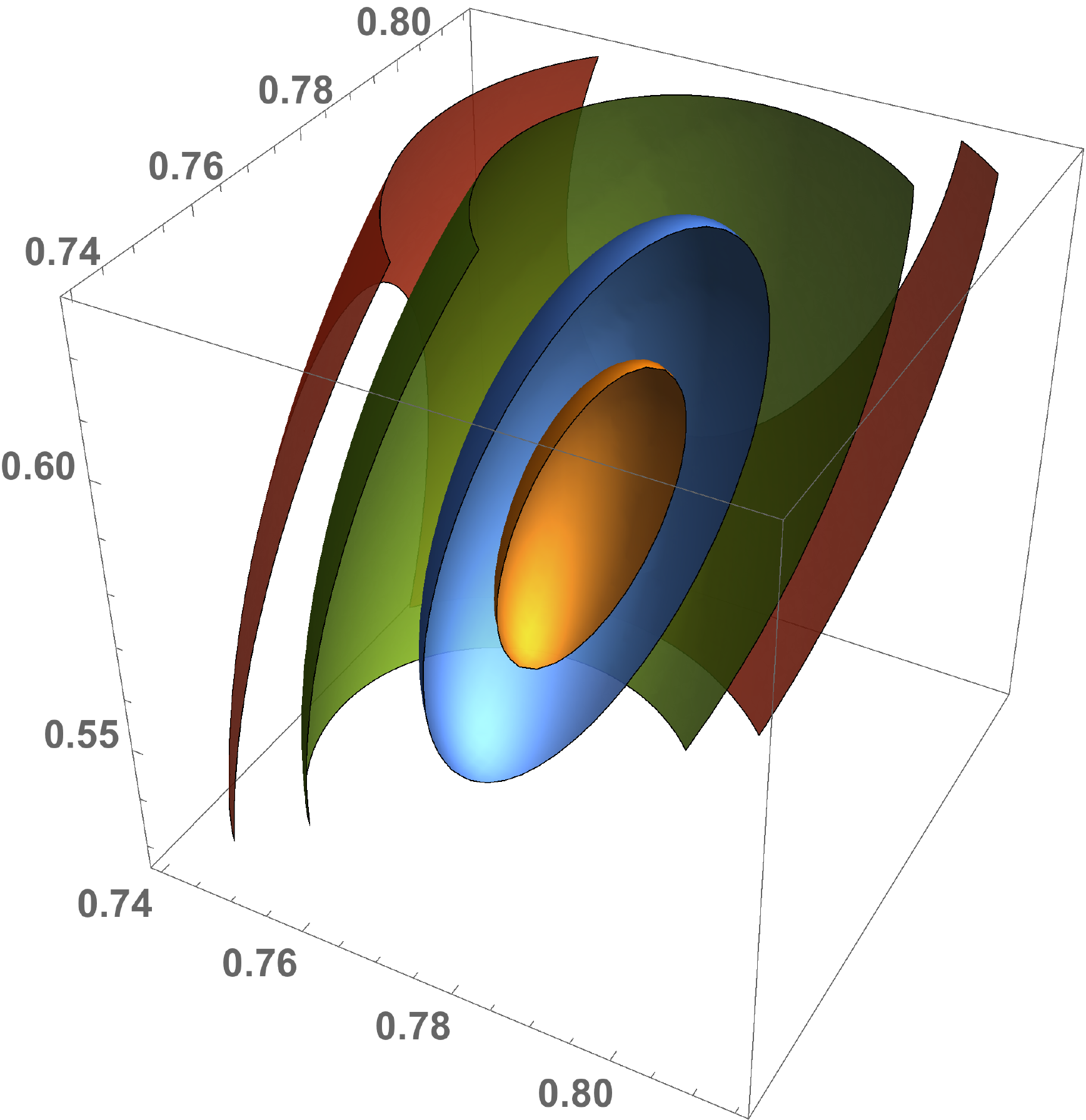}
\begin{picture}(0.1,0.1)(0,0)
\put(-220,105){\makebox(0,0){\footnotesize{$R$}}}
\put(-180,210){\makebox(0,0){\footnotesize{$\hat k_y$}}}
\put(-150,10){\makebox(0,0){\footnotesize{$\hat k_x$}}}
\end{picture}
\caption{Surfaces of constant free energy density $\hat{w}\equiv w/\mu^{3}$ in the three dimensional module space $(\hat{k}_{x},\hat{k}_{y},R_0)$, where $\hat{k}_{x}\equiv k_{x}/\mu$ and $\hat{k}_{y}=k_{y}/\mu$. The plot was made for the $T/\mu=0.04$ case of figure \ref{fig:fingerprint}. The minimum is for the triangular lattice with $R=1/\sqrt{3}$ and equal periods $\hat{k}_{x}=\hat{k}_{y},=0.777$.
\label{fig:3DFE}}
\end{figure}

To explore whether the triangular lattice is preferred in the full three dimensional moduli space, we also
constructed additional black holes, about 600, at the lowest temperature of figure \ref{fig:fingerprint}, $T/\mu=0.04$, in a neighbourhood 
of the triangular lattices. In figure \ref{fig:3DFE} we see the triangular lattice is indeed a local minimum of the free energy
and almost certainly the global minimum.

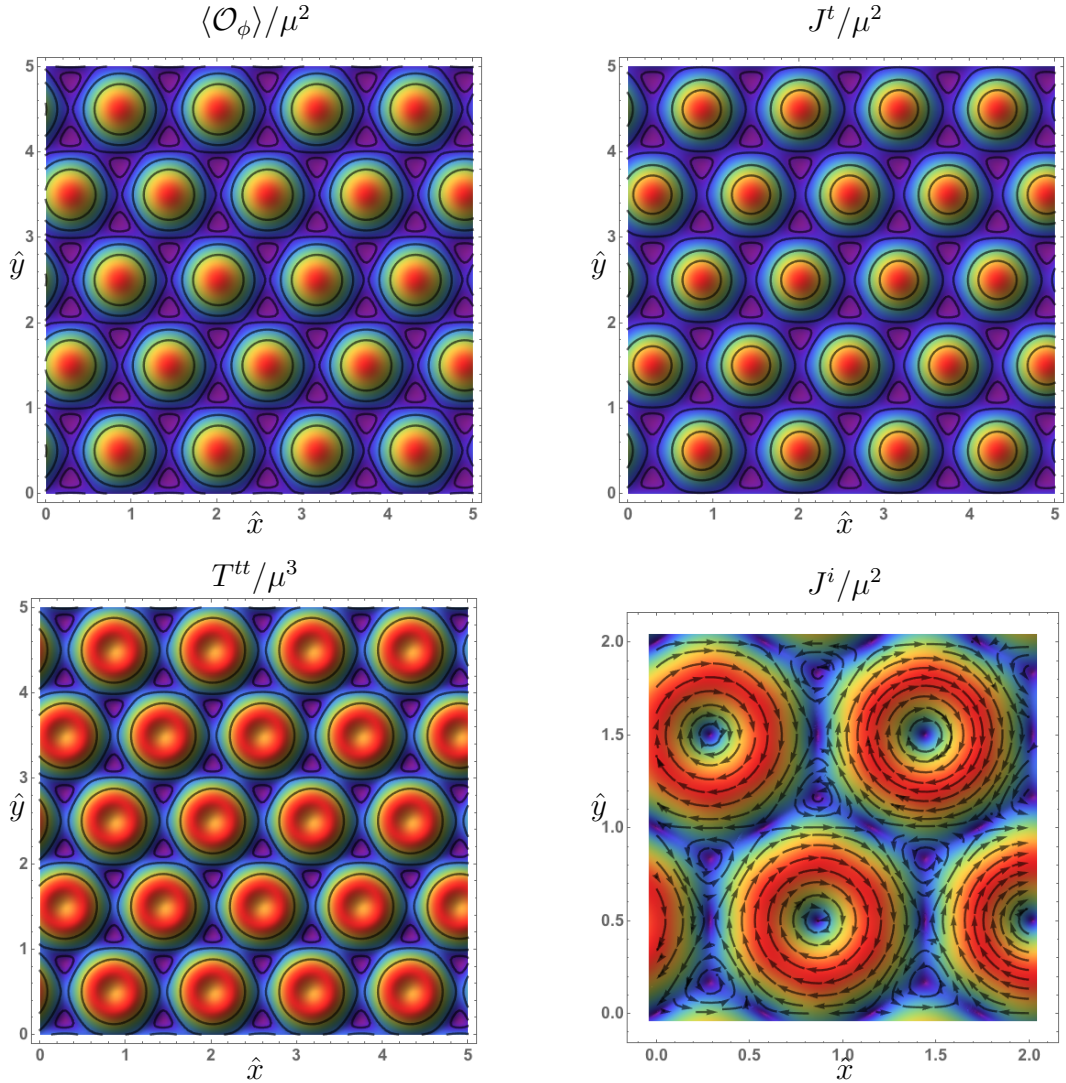
\begin{figure}[h!]
\centering
\includegraphics[width=0.4\textwidth]{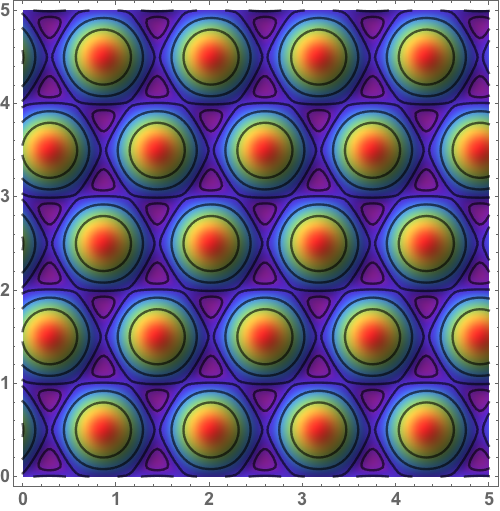}\qquad\qquad
\includegraphics[width=0.4\textwidth]{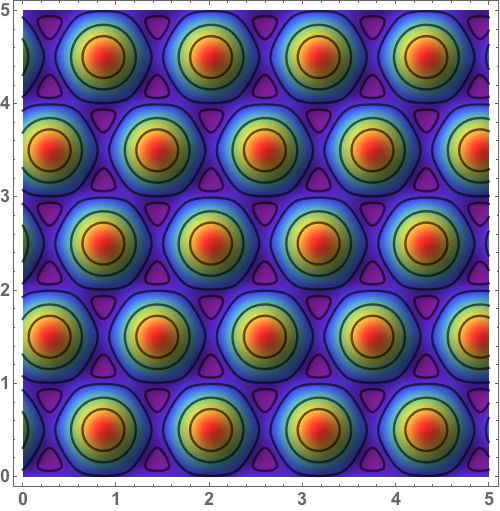}\\
\vspace{27pt}
\includegraphics[width=0.4\textwidth]{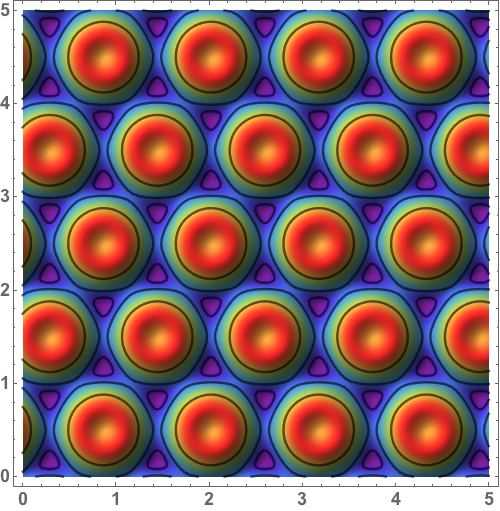}\qquad\qquad
\includegraphics[width=0.4\textwidth]{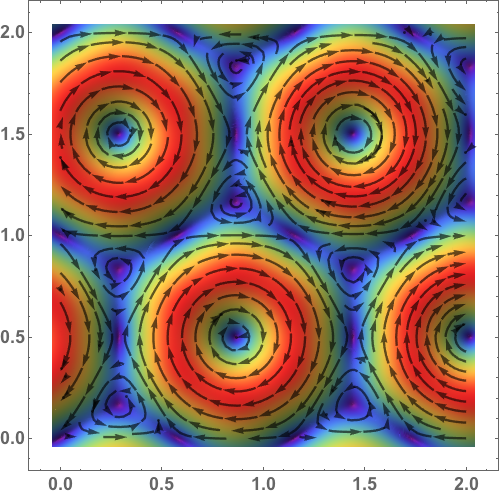}
\begin{picture}(0.1,0.1)(0,0)
\put(-398,303){\makebox(0,0){{$\hat y$}}}
\put(-178,303){\makebox(0,0){{$\hat y$}}}
\put(-398,98){\makebox(0,0){{$\hat y$}}}
\put(-178,98){\makebox(0,0){{$\hat y$}}}
\put(-308,206){\makebox(0,0){{$\hat x$}}}
\put(-85,206){\makebox(0,0){{$\hat x$}}}
\put(-308,-1){\makebox(0,0){{$\hat x$}}}
\put(-85,-1){\makebox(0,0){{$\hat x$}}}
\put(-308,395){\makebox(0,0){{$\langle\mathcal{O}_\phi\rangle/\mu^2$}}}
\put(-85,395){\makebox(0,0){{$J^t/\mu^2$}}}
\put(-308,185){\makebox(0,0){{$T^{tt}/\mu^3$}}}
\put(-85,182){\makebox(0,0){{$J^i/\mu^2$}}}
\end{picture}
\caption{The spatial behaviour of various expectation values for the thermodynamically preferred triangular lattice at $T/\mu=0.04$.
Top left shows the expectation value of the pseudo scalar operator 
$\langle\mathcal{O}_\phi\rangle/\mu^2$, top right shows the charge density $J^t/\mu^2$ ,
bottom left shows $T^{tt}/\mu^3$ and bottom right shows the flow lines and the norm of the magnetisation currents
$J^i/\mu^2$. The plots are functions
of the spatial coordinates $(\hat x,\hat y)\equiv (\frac{x'}{L_x},\frac{y'}{L_y})$ and
colours with longer wavelength correspond to larger values (the norm of $J^i/\mu^2$ for the bottom right).
\label{honey}}
\end{figure}

It is interesting to display the spatial variation of various expectation values for the thermodynamically preferred triangular lattices.
The general features are the same for all temperatures that we have considered.
In figure \ref{honey}, for the representative case of $T/\mu=0.04$, we display 
the spatial variation of the operator $\mathcal{O}_\phi$, dual to the 
pseudoscalar field, the charge density, $J^t$, the energy density, $T^{tt}$, and also the norm and direction of
the spatial components
of the current vector $J^i$. It is interesting to notice that $\mathcal{O}_\phi$ and $J^t$ have a spiky structure
and $T^{tt}$ and $J^i$ have dimples at the positions of the spikes.

\begin{figure}[t!]
\centering
\includegraphics[width=0.47\textwidth]{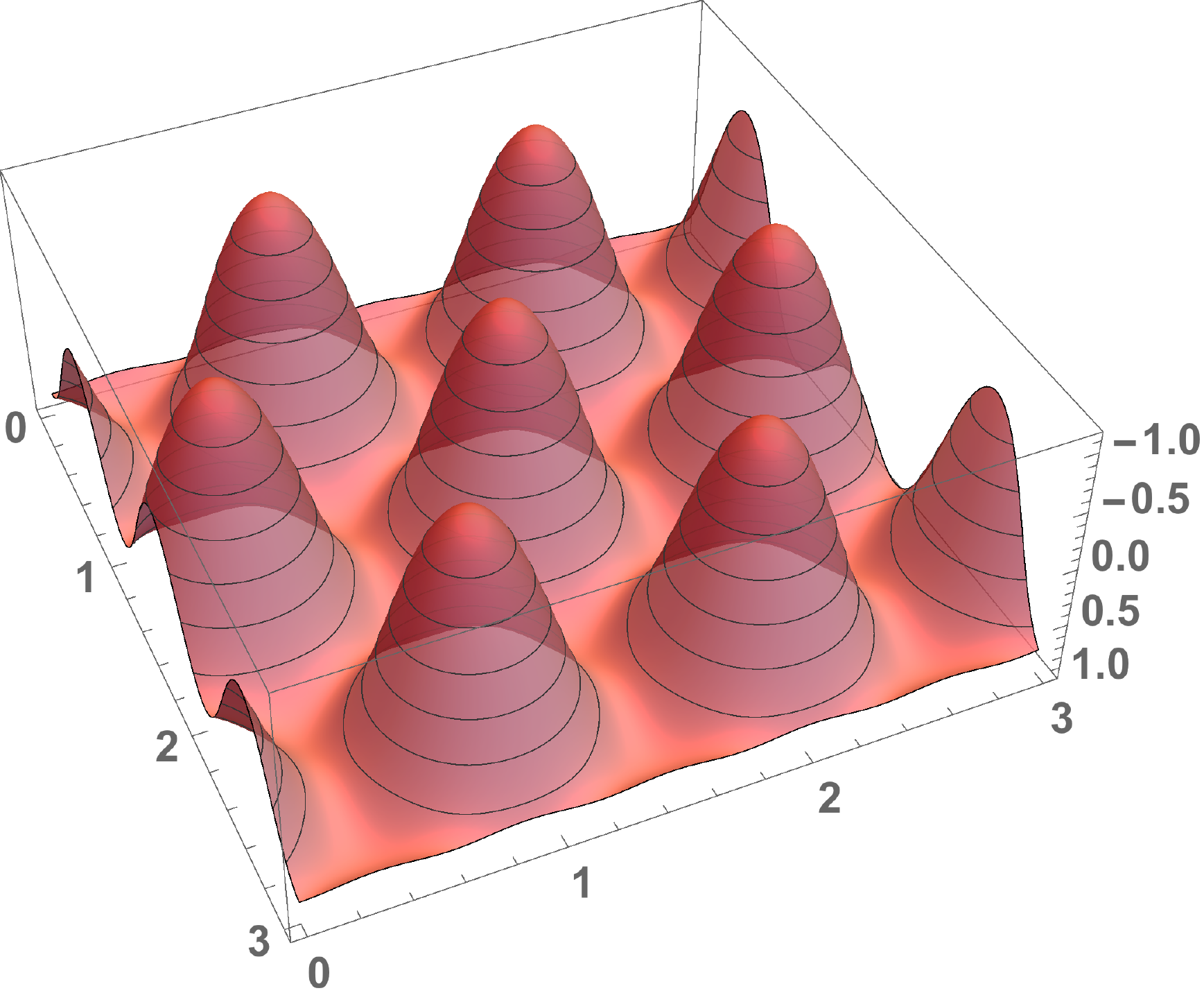}\qquad
\includegraphics[width=0.47\textwidth]{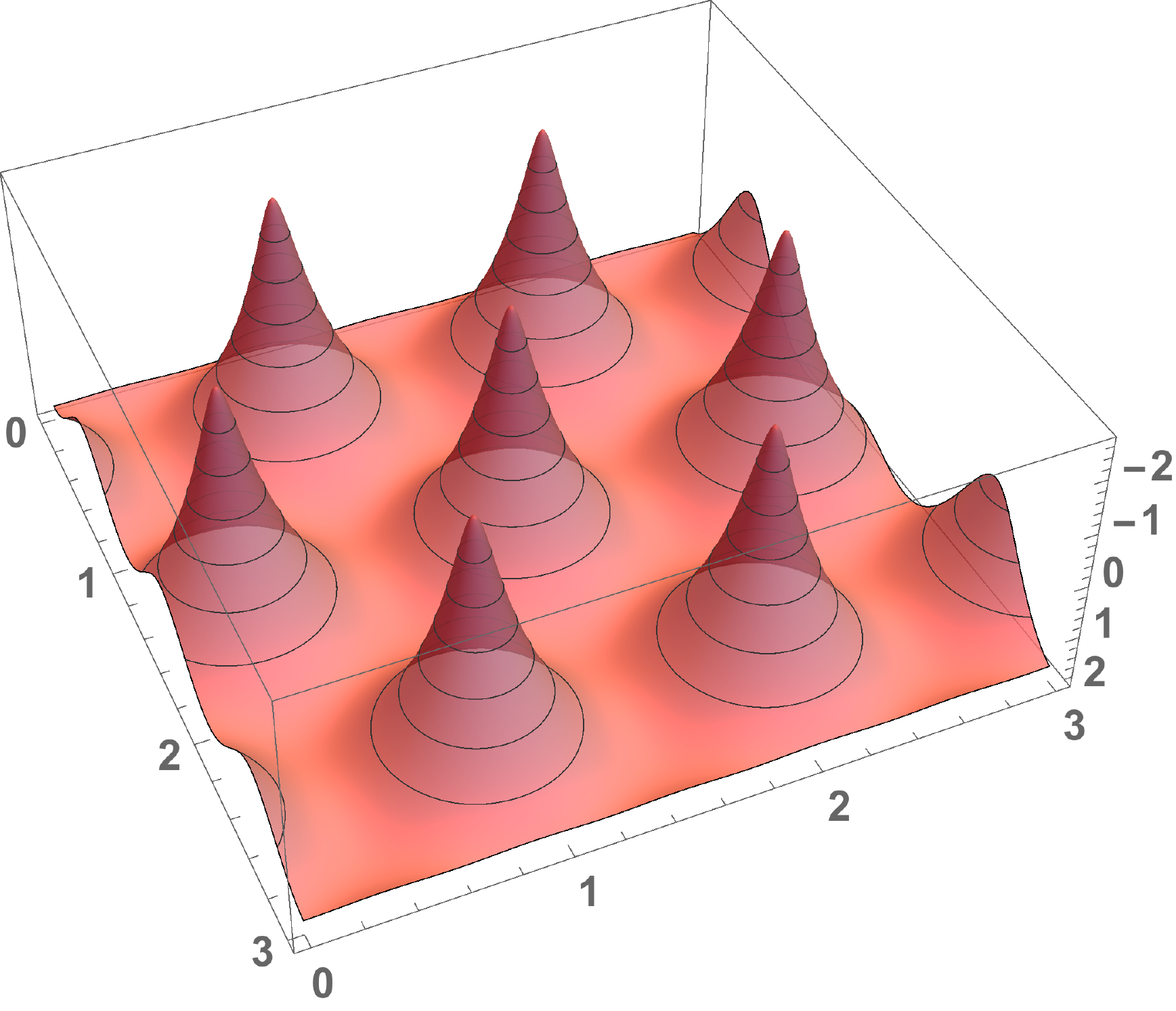}
\begin{picture}(0.1,0.1)(0,0)
\put(-85,35){\makebox(0,0){{$\hat x$}}}
\put(140,35){\makebox(0,0){{$\hat x$}}}
\put(-205,70){\makebox(0,0){{$\hat y$}}}
\put(25,70){\makebox(0,0){{$\hat y$}}}
\put(-5,90){\makebox(0,0){{$\phi_+$}}}
\put(225,90){\makebox(0,0){{$\phi_+$}}}
\end{picture}
\caption{A plot of the spatial dependence of the scalar field on the black hole horizon, $\phi_+\equiv\phi(r_+)$,
as a function of the spatial coordinates $(\hat x,\hat y)\equiv (\frac{x'}{L_x},\frac{y'}{L_y})$. The left plot is
for the thermodynamically preferred triangular lattice at $T/\mu=0.04$. The right plot is for the preferred
triangular lattice at $T/\mu=0.01$.
\label{spikyhorizon}}
\end{figure}
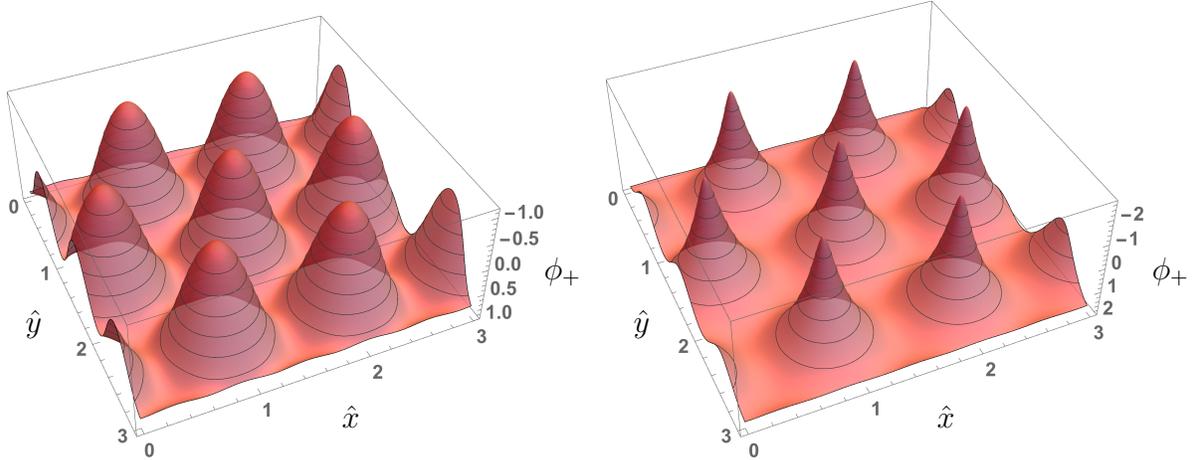

It would be very interesting to determine the ultimate $T=0$ ground state of the spatially modulated phase.
It seems likely that it continues to be a triangular lattice for all temperatures. Following the argument
in section 3.7.5 of \cite{Donos:2013woa} we expect that $\partial_Tk_x\to 0$ as $T\to 0$
and it is plausible this happens at $k_x\ne 0$, leading to a crystalline ground state. 
It is numerically challenging to keep reducing the temperature 
in the systematic manner that we have described so far. In order to get some additional insight into the ground state,
we constructed the thermodynamically preferred triangular lattice at $T/\mu=0.01$, i.e. amongst the
triangular lattices we determined the thermodynamically preferred value of the wavenumber, finding
$k/\mu=0.6456$. 
In figure \ref{spikyhorizon} we have plotted the spatial variation of the pseudoscalar field on the black hole horizon
for both this temperature and also for $T/\mu=0.04$. The appearance of spikes at the horizon 
as we lower the temperature, combined with the structure of the magnetisation current displayed in 
figure \ref{honey},
indicates that the ground state is a fundamentally new type of holographic solution, which will be studied in more detail in the future.

%\begin{figure}[t!]
%\centering
%\includegraphics[width=0.47\textwidth]{JStreamPlot.png}
%\caption{A plot of the integral curves of the abelian current density $J^i$, 
%as a function of the spatial coordinates $(\hat x,\hat y)\equiv (\frac{x'}{L_x},\frac{y'}{L_y})$.  Hotter colours
%correspond to current vectors with larger lengths.
%The plot is
%for the thermodynamically preferred triangular lattice at $T/\mu=0.01$.
%\label{magcur}}
%\end{figure}

\section{Discussion}
In this paper we have numerically constructed co-homogeneity three, asymptotically AdS black holes in
four spacetime dimensions. The solutions are holographically dual to $d=3$ CFTs
at finite chemical potential and in a constant magnetic field, which spontaneously form
a periodic lattice in two spatial dimensions, with magnetisation currents
of the form \eqref{defjays} circulating the plaquettes of the lattice. In addition there is also a commensurate modulation
of the charge density. These features, which were also observed in the checkerboard lattice of \cite{Withers:2014sja}, 
are somewhat reminiscent of the current loop order that is observed in the cuprates \cite{vojta}.

We showed that the black holes come in three-parameter families which are associated with
different Bravais lattices. For a specific value of the magnetic field we showed
that the triangular lattice is thermodynamically preferred over all other lattices, at least down to very low temperatures.
While this is certainly a very natural lattice to appear, being associated with close packing of circles in the plane,
it is unclear what aspects of the gravitational model selects this lattice. Perhaps it is possible\footnote{Difficulties
in relating Landau-Ginzburg theory to holography were recently discussed in \cite{Banks:2015aca}.}
to illuminate this issue by developing a Landau-Ginzburg type description near the critical temperature. However,
such a description will not be valid at low temperatures. It would be interesting to consider other values of 
the magnetic field and see if the triangular lattice persists. It would be particularly interesting to examine what happens
as we reduce the magnetic field to zero since, based on the work of \cite{Withers:2014sja},
it seems likely that the striped phase is then the preferred configuration. 
More generally, it is essentially a wide open question which periodic lattices in two spatial dimensions can be preferred
in the context of more general holographic constructions. The possibilities in three spatial
dimensions, associated with five dimensional gravitational models, are richer still. 
%We continue to be impressed with the
%extraordinarily diverse landscape of asymptotically AdS black holes that exist.

It would be interesting to examine the transport properties of the new triangular phase that we have found. 
Powerful techniques have been recently developed that enable one to extract the thermoelectric
DC conductivity from the black hole horizon \cite{Donos:2015gia,Banks:2015wha,Donos:2015bxe}. 
However, since the triangular phase spontaneously breaks translations, 
there will be Goldstone modes and hence the DC conductivity is expected to be infinite. 
On the other hand the intricate structure of magnetisation
currents in our phase could lead to interesting structures in the AC conductivity. It should be noted, however, 
that while it is conceptually straightforward to calculate the thermoelectric 
conductivity by perturbing the black holes, it is technically somewhat involved.

The low temperature limit of the black hole solutions that we constructed also revealed some striking features. 
The spatial modulation of the triangular lattice does not got washed out, but instead persists.
Moreover, the spatial modulation at the black hole horizon starts to develop a spiky structure which
certainly deserves to be explored in more detail. Indeed it suggests that the zero temperature
ground state is a triangular crystal that is being supported by a novel periodic structure on the horizon.

\section*{Acknowledgements}

We would like to thank Tom Griffin, Luis Melgar and Toby Wiseman for helpful discussions. 
Part of this work was undertaken on the COSMOS Shared Memory system at DAMTP, University of Cambridge operated on behalf of the STFC DiRAC HPC Facility. This equipment is funded by BIS National E-infrastructure capital grant ST/J005673/1 and STFC grants ST/H008586/1, ST/K00333X/1. A significant part of our computation were carried out at Durham University's Hamilton HPC cluster.
The work is also supported by STFC grant ST/J0003533/1, EPSRC grant EP/K034456/1,
and by the European Research Council under the European Union's Seventh Framework Programme (FP7/2007-2013), ERC Grant agreements STG 279943 and ADG 339140.
\appendix

\section{Asymptotic expansion of fields}\label{app:expansion}
Using the modified equations of motion described in section \ref{sec:non-linear}, the 
expansion of the functions appearing in the ansatz \eqref{eq:pde_ansatz}
near the AdS boundary have the form
\begin{align}\label{asg1g2exp}
Q_{tt}\left(r,x,y\right)&=1+r^{3}\,c_{tt}(x,y)+g_{1}(x,y)\,r^{(3+\sqrt{33})/2}+\mathcal{O}(r^{4})\,,\nn
Q_{rr}\left(r,x,y\right)&=1+g_{2}(x,y)\,r^{(3+\sqrt{33})/2}+\mathcal{O}(r^{4})\,,\nn
Q\left(r,x,y\right)&=1-\frac{1}{2}\,r^{3}\,c_{tt}(x,y)+g_{1}(x,y)\,r^{(3+\sqrt{33})/2}+\mathcal{O}(r^{4})\,,\nn
B\left(r,x,y\right)&=1+r^{3}\,c_{B}(x,y)+\mathcal{O}(r^{4}),\quad
R\left(r,x,y\right)=R_{0}+r^{3}\,c_{R}(x,y)+\mathcal{O}(r^{4})\,,\nn
Q_{tr}\left(r,x,y\right)&=r^{4}\,c_{tr}(x,y)+\mathcal{O}(r^{4}\,\ln r),\quad
Q_{tx}\left(r,x,y\right)=r^{3}\,c_{tx}(x,y)+\mathcal{O}(r^{4})\,,\nn
Q_{ty}\left(r,x,y\right)&=r^{3}\,c_{ty}(x,y)+\mathcal{O}(r^{4}),\quad
Q_{rx}\left(r,x,y\right)=r^{4}\,c_{rx}(x,y)+\mathcal{O}(r^{4}\,\ln r)\,,\nn
Q_{ry}\left(r,x,y\right)&=r^{4}\,c_{ry}(x,y)+\mathcal{O}(r^{4}\,\ln r),\quad
a_{t}\left(r,x,y\right)=\mu+r\,c_{t}(x,y)+\mathcal{O}(r^{2})\,,\nn
a_{r}\left(r,x,y\right)&=r^{3}\,c_{r}(x,y)+\mathcal{O}(r^{3}\,\ln r),\quad
a_{x}\left(r,x,y\right)=r\,c_{x}(x,y)+\mathcal{O}(r^{2})\,,\nn
a_{y}\left(r,x,y\right)&=r\,c_{y}(x,y)+\mathcal{O}(r^{2}),\quad
\varphi\left(r,x,y\right)=r\,c_{\varphi}(x,y)+\mathcal{O}(r^{2})\,.
\end{align}
In total we have fifteen functions of $x$ and $y$ which are fixed by integration. Analogues of the
functions $g_{1}$ and $g_{2}$ first appeared in a closely related context in \cite{Donos:2014yya}.
The condition $\xi^\mu=0$ implies a linear relation between $g_1$ and $g_2$ and they can then be removed after
doing a coordinate transformation. Similarly, the functions $c_{tr}$, $c_{xr}$, $c_{yr}$ and $c_r$ can also
be removed by coordinate and gauge transformation. In particular, these functions do not appear in
the expressions for physical expectation values of the dual field theory.

The components of the abelian current vector can be expressed in terms of the above expansion via
\begin{align}
J^{t}&=r_{+}\,\left(\mu-\frac{1}{2}\,c_{t} \right)\,,\nn
J^{x}&=\frac{r_{+}}{2\hat{L}_{x}}\,\left(\sqrt{1+R_{0}^{2}}\,c_{x}-R_{0}\,c_{y} \right)\,,\nn
J^{y}&=\frac{r_{+}}{2\hat{L}_{y}}\,\left(\sqrt{1+R_{0}^{2}}\,c_{y}-R_{0}\,c_{x} \right)\,.
\end{align}
Similarly, the components of the stress energy tensor are given by
\begin{align}\label{eq:stress_tensor}
T^{t}{}_{t}&=-\frac{B^{2}}{2r_{+}}-\frac{\mu^{2}r_{+}}{2}-2\,r_{+}^{3}+\frac{3}{4}r_{+}^{3}\,c_{tt}\,,\nn
T^{x}{}_{x}&=\frac{B^{2}}{4r_{+}}+\frac{\mu^{2}r_{+}}{4}+r_{+}^{3}-\frac{3}{8}r_{+}^{3}\,c_{tt}+\frac{3}{4}r_+^3\,\left(1+R_{0}^{2}\right)\,c_{B}\,,\nn
T^{y}{}_{y}&=\frac{B^{2}}{4r_{+}}+\frac{\mu^{2}r_{+}}{4}+r_{+}^{3}-\frac{3}{8}r_{+}^{3}\,c_{tt}-\frac{3}{4}r_+^3\,\left(1+R_{0}^{2}\right)\,c_{B}\,,\nn
T^{x}{}_{y}&=\frac{3L_{y}r_{+}^{3}}{4L_{x}\sqrt{1+R_{0}^{2}}}\,\left(R_{0}\left(1+R_{0}^{2}\right)\,c_{B}+c_{R} \right)\,,\nn
T^{t}{}_{x}&=\frac{3}{4}\hat{L}_{x}\,r_{+}^{3}\,c_{tx},\quad T^{t}{}_{y}=\frac{3}{4}\hat{L}_{x}\,r_{+}^{3}\,c_{ty}\,.
\end{align}
In obtaining these results we note that to get the boundary metric in the standard form \eqref{bc1} one needs to scale
$r=r_+\epsilon/2$. 
Note that for the thermodynamically preferred black holes, satisfying \eqref{extreme},
we should have 
\begin{align}
\bar c_B=\bar c_R=0
\end{align}
which we have checked in our numerics.

\subsection{Magnetisation currents}\label{magcurr}
Following \cite{Donos:2015bxe} we show that the spatial components of the 
abelian and momentum currents in the dual CFT
are magnetisation currents and hence the average current fluxes vanish: $\bar J^{i}=\bar T^{ti}=0$.

Consider the {\it bulk} current defined via
\begin{align}
J^a_{bulk}=\sqrt{-g}\left[Z{F}^{ar}+\frac{1}{2}\vartheta\epsilon^{ar\rho_1\rho_2}F_{\rho_1\rho_2}\right]\,,
\end{align}
and observe that when evaluated at the AdS boundary $J^a_{bulk}$ is the conserved
current density, $\sqrt{\gamma}J^a$, of the dual CFT.
We then consider the radial derivative, $\partial_r J^i_{bulk}$, use the gauge equation of motion on the right hand side and then integrate over $r$ from the horizon to
the asymptotic boundary. After noticing that $J^i_{bulk}$ vanishes at the horizon we deduce that
the spatial part of current density of the dual field theory is a magnetisation current:
\begin{align}\label{ccon}
\sqrt{\gamma}J^i=\partial_j M^{ij}\,,
\end{align}
where the local thermal magnetisation density, $M^{ij}$, is given by
\begin{align}\label{defM}
M^{ij}=-\int_0^\infty dr \sqrt{-g}\left( ZF^{ij}+\vartheta\epsilon^{ijtr}F_{tr}\right)\,.
\end{align}
In fact this procedure only defines $M^{ij}$ up to a constant times $\epsilon^{ij}$. We have fixed
this constant in \eqref{defM} by demanding that the zero mode gives rise to the magnetisation
density $m$, given in \eqref{defemm}, and defined by the first law \eqref{varfe}.
We observe that \eqref{ccon} implies $\bar J^i$=0. 
Thus, in the expansion \eqref{asg1g2exp}, by considering \eqref{eq:stress_tensor},
we must have $\bar c_{tx}=\bar c_{ty}=0$ for all black hole solutions, and we have checked this in our numerics.

A similar story unfolds for the heat current.
We first define 
\begin{align}\label{defgee}
G^{\mu\nu}\equiv-2\nabla^{[\mu}k^{\nu]}-{Z}k^{[\mu}{F}^{\nu]\sigma}A_\sigma-\frac{1}{2}(\varphi-\theta)
\left[Z{F}^{\mu\nu}
+\frac{1}{2}\vartheta\epsilon^{\mu\nu\rho_1\rho_2}F_{\rho_1\rho_2}\right]\,,
\end{align}
where $\varphi\equiv i_kA$ and we write $i_kF\equiv \psi+d\theta$ for a globally defined function $\theta$. We
will take $\theta=-A_t$, so that $\varphi=-\theta=A_t$ and $\psi_\nu=\partial_t A_\nu$.
We find
\begin{align}
\nabla_\mu G^{\mu\nu}
&=\left({V}-\frac{1}{8}A_\sigma\nabla_\mu\vartheta\epsilon^{\mu\sigma\rho_1\rho_2}F_{\rho_1\rho_2}  \right)k^\nu+\frac{1}{2}Z{F}^{\nu\mu}\psi_\mu\nn
&\qquad\qquad\qquad-\frac{1}{2}{ZA_\sigma\mathcal{L}_k({F}^{\nu\sigma})}
-\frac{1}{4}\nabla_\mu(\varphi-\theta)\vartheta\epsilon^{\mu\nu\rho_1\rho_2}F_{\rho_1\rho_2}\,.\label{Gder}
\end{align}
We now define
\begin{align}
Q^a_{bulk}\equiv\sqrt{-g}G^{ar}\,,
\end{align}
and observe that when evaluated at the AdS boundary $Q^a_{bulk}$ is the heat current density, $\sqrt{\gamma}(T^{ta}-\mu J^a)$, of the dual CFT. A 
calculation similar to above implies
\begin{align}\label{jayback}
\sqrt{\gamma}(T^{ti}-\mu J^i)&=\partial_j M^{ij}_T\,,
\end{align}
where the local thermal magnetisation density,
$M^{ij}_T(x)$, is given by
\begin{align}\label{emmslocal}
M_T^{ij}&=-\int_0^\infty dr\sqrt{-g}G^{ij}\,.
\end{align}
In particular, since $\bar J^{i}=0$, we can conclude that $\bar T^{ti}=0$ also.
Hence, in the expansion \eqref{asg1g2exp}, by considering \eqref{eq:stress_tensor},
we deduce that $\bar c_{tx}=\bar c_{ty}=0$ for all black hole solutions, and we have checked this in our numerics.

\section{Numerical convergence}\label{app:reference}
In order to solve the system of PDEs with the boundary conditions explained in section \ref{sec:non-linear} we discretise the computational domain $\left[0,1\right]\times \left[0,1\right]\times\left[0,1\right]$ on $N_{r}$, $N_{x}$ and $N_{y}$ points respectively. As we will describe in the next paragraph, after approximating the partial derivatives on these points by an interpolation method, we are solving the resulting algebraic system of equations by using Newton's method.

Since our boundary conditions are periodic in the $x$ and $y$ directions, we use a Fourier spectral method in order to approximate the partial derivatives in these directions. This amounts to placing our points at equal distances $1/N_{x}$ and $1/N_{y}$ in those directions. For the discretisation in the ``radial'' coordinate $r$ we have tested both Chebyshev spectral method and finite differences.

\begin{figure}[t!]
\centering
\includegraphics[width=0.47\textwidth]{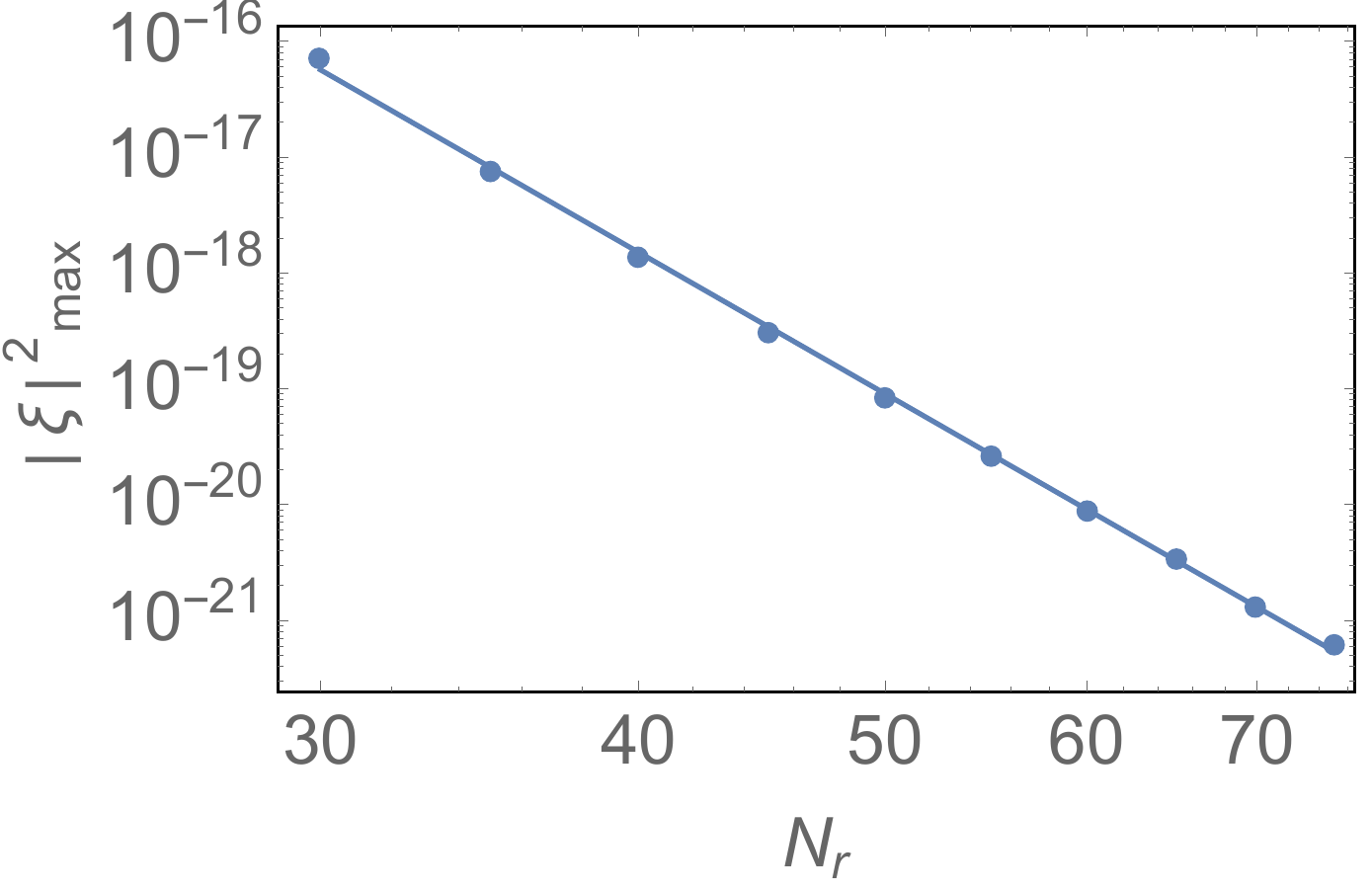}\qquad
\includegraphics[width=0.47\textwidth]{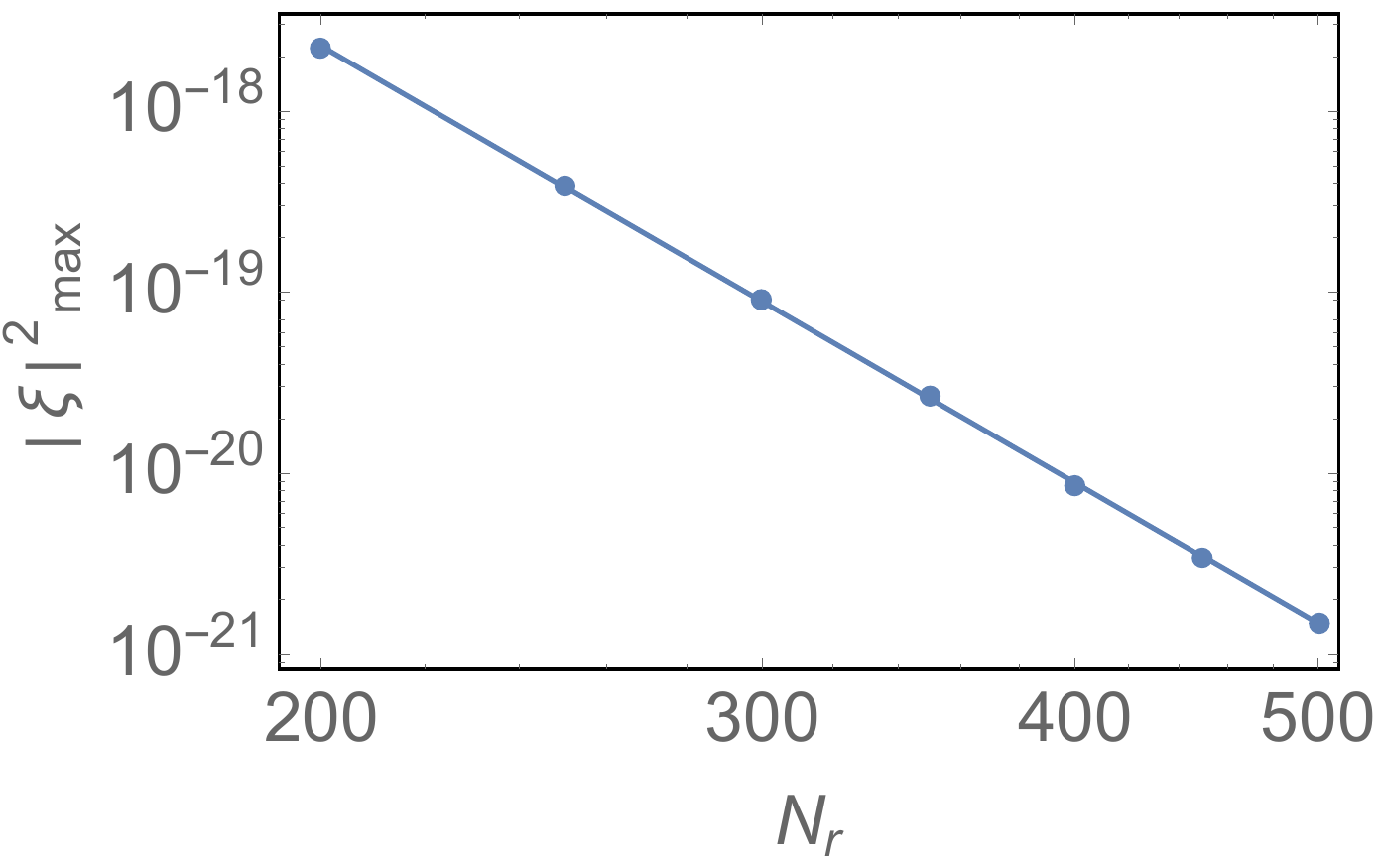}
\caption{Convergence plots of the maximum value of $\left|\xi \right|^{2}_{max}$ in our computational grid as a function of grid points in the radial direction $r$ for Chebyshev spectral (left) and fourth order finite differences (right). Both plots are for $T/\mu=0.05$, $k_{x}=k_{y}=0.97\,\mu$, $R_{0}=0$ and $B=0.05\,\mu^{2}$.
\label{fig:convergence}}
\end{figure}

In figure \ref{fig:convergence} we present two convergence plots for the maximum value of $\left|\xi \right|^{2}_{max}$ in our computational grid for a representative lattice (in this case a checkerboard lattice). 
We have used $N_{x}=N_{y}=60$ points in the periodic directions for both cases of spectral and finite difference grids in the radial directions. For both both cases we have found power law convergence, $\left|\xi \right|^{2}_{max}\sim N^{-12.6}$ for spectral and $\left|\xi \right|^{2}_{max}\sim N^{-8.01}$ for fourth order finite differences. We can understand \cite{Donos:2014yya} the lack of 
spectral convergence (i.e. exponential convergence) from the non-analytic field expansion close to boundary
(see \eqref{asg1g2exp}). We have also checked that $\psi$, defined
in \eqref{defpsi}, satisfisfies $\psi<10^{-10}$ for all of the solutions we have constructed.

We have implemented our numerical method in \verb!C++!. The facility of class templates has been particularly helpful to accommodate the various numerical precisions we have used at low temperatures and in the convergence tests. At certain key points of our code we have specialised our templates to double, long double, and \verb!__float128!. In order to generate and manipulate our data resulting from the different points of our domain to different computing nodes, we have followed a hybrid approach to parallelisation using a combination of \verb!MPI! and \verb!OpenMP!. After obtaining the matrix of variations in Newton's method implementation, we used the library \verb!PETSc! \cite{petsc-web-page} in order to solve the resulting linear system of equations. In order to achieve that, we used a block-ILU(0) preconditioner in combination with a GMRES iterative method. The typical system we had to solve was a $\sim 10^{6}\times 10^{6}$ system with $\sim 10^{10}$ non-zero matrix elements for the matrix we needed to invert.

%\bibliographystyle{utphys}
%\bibliography{helical}{}
%\end{document}

\providecommand{\href}[2]{#2}\begingroup\raggedright\endgroup

\end{document}